\documentclass[nofootinbib,tightenlines,amsmath,amssymb,aps]{revtex4}

\newcommand{\beq}{\begin{equation}}
\newcommand{\eeq}{\end{equation}}
\newcommand{\be}{\begin{equation}}
\newcommand{\ee}{\end{equation}}
\newcommand{\beqa}{\begin{eqnarray}}
\newcommand{\eeqa}{\end{eqnarray}}

\newcommand{\bean}{\begin{eqnarray*}}
\newcommand{\eean}{\end{eqnarray*}}
\newcommand{\RR}{\mathbb{R}}
\newcommand{\ZZ}{\mathbb{Z}}
\newcommand{\extd}{{\mathrm {d}}}

\newcommand{\pa}{\partial}

\newcommand{\OO}{\mathcal{O}_{\Gamma}}
\newcommand{\VV}{\mathcal{V}}
\newcommand{\A}{\mathcal{A}}
\newcommand{\SSS}{\mathcal{S}}

\newcommand{\s}{\sigma}
\newcommand{\e}{\epsilon}

\usepackage{psfrag}
\usepackage[dvips]{graphicx}

\begin{document}

\title{
{\center
Hidden Quantum Gravity in 4d Feynman diagrams:\\
 Emergence of spin foams}
}

\author{Aristide Baratin$^{1,2}$\thanks{email: abaratin@perimeterinstitute.ca},
Laurent Freidel$^{1,2}$\thanks{email:
lfreidel@perimeterinstitute.ca}}

\affiliation{\centerline{\footnotesize \it
$^1$Laboratoire de Physique, \'Ecole Normale Sup{\'e}rieure de Lyon} \\
\centerline{\footnotesize \it 46 all{\'e}e d'Italie, 69364 Lyon
Cedex 07, France.}\\
\centerline{\footnotesize \it $^2$Perimeter Institute for
Theoretical Physics} \\
 \centerline{\footnotesize \it 31 Caroline street North, N2L2Y5
Waterloo, ON, Canada}}

\begin{abstract}
We show how Feynman amplitudes of standard QFT on flat and
homogeneous space can naturally be recast as the evaluation of
observables for a specific spin foam model, which provides
dynamics for the background geometry. We identify the symmetries
of this Feynman graph spin foam model and give the gauge-fixing
prescriptions. We also show that the gauge-fixed partition
function is invariant under Pachner moves of the triangulation,
and thus defines an invariant of four-dimensional manifolds.
Finally, we investigate the algebraic structure of the model, and
discuss its relation with a quantization of 4d gravity in the
limit $G_N \to 0$.

\end{abstract} \maketitle

\tableofcontents

\newpage

%%%%%%%%%%%%%%%%%%%%%%%%%%%%%%%%%%%%%%%%%%%%%%%%%
\section{Introduction}
%%%%%%%%%%%%%%%%%%%%%%%%%%%%%%%%%%%%%%%%%%%%%%%%%%

One of the main challenges faced  by anybody thinking  seriously
about quantum gravity is the fact that such a theory should be
understood and formulated in a background independent manner. That
is, classical space-time should emerge as a low energy approximation
of a more fundamental - and yet unknown - description of quantum
spacetime; and a vacuum selection principle should be designed in
order to give a dynamical understanding for the emergence of the
particular type of space-time we live in among all possibilities.
It also means that one should be able to have a proper handle on
the observables of quantum gravity, which are intrinsically non
local because of background independence - or more precisely
diffeomorphism invariance. This basic and fundamental challenge is
however at odds with our well established current understanding of
fundamental physics formulated in terms of local quantum field
theory living on a fixed background. This schizophrenic state of
affair seems to force a painful choice between the questions we
want to address and the fundamental techniques at our disposal.

If one takes the point of view that understanding quantum gravity in
a background independent manner is the key to success, one is led
to first devise a new set of appropriate tools and techniques well
tailored to this problem. There has been, in the recent years, a
large body of work in that direction. Such works have led to the
conclusion, or more appropriately the hypothesis, that the proper
tools are, at the kinematical level, given by spin networks as
developed in loop quantum gravity; and at the dynamical level, given
by the so-called spin foam models. Spin foam models give a well
defined framework allowing to address the dynamical problem of
quantizing gravity in a background independent manner, and provide
a description of quantum space-times in a purely algebraic and
combinatorial way \cite{review}. The state of development is such that one can
now propose, for Euclidean 4-dimensional pure
gravity, well defined and finite quantum gravity transition
amplitudes, which are independent of any triangulation or
undesirable discrete structure \cite{Lgft}.

Is it a satisfying state of affair? For a specialist working along
this line of thought, there are many reasons to be satisfied with
all the new developments; however the answer is clearly no. The
answer is negative since it is not yet possible to convincingly
argue that, when the Newton constant $G_{N}$ is treated as a small
parameter, this set of amplitudes reproduces local quantum field
theory; or that, when $\hbar$ is treated as a small parameter, one
recovers the dynamics of general relativity.

Some recent progress have been achieved recently concerning the later
problem in the context of spin foam \cite{Carloprop} and in the Hamiltonian framework \cite{Thomas},
but still more work is needed.
%The idea in this work is to compute an overlap between the spin foam amplitude and a semi-classical coherent state in order to extract
%the graviton propagator from these models. This can be successfully done in a restricted class of configuration,
%one of the challenge being to extend this new strategy to general configurations.

This means that we
cannot yet falsify the spin foam hypothesis; it is for us a
serious shortcoming of all the developments in background
independent approach to quantum gravity, since any physical theory
should expose itself to falsifiability tests.

These serious problems are related to the fact that it is
extremely hard, if not impossible \cite{Torre}, to construct proper
observables having a clear physical meaning in the context of background independent pure gravity without
matter fields\footnote{Note that recently a new proposal has been made to overcome this difficulty\cite{Carloprop}.
The idea in this work is to compute an overlap between the spin foam amplitude and a semi-classical coherent state in order to extract
the graviton propagator from these models. This can be successfully done in a restricted class of configurations,
one of the challenge being to extend this new strategy to general configurations.}.
This is not really surprising, given that one can argue that space-time
geometry is a mathematical abstraction devised to account in a
simple way for all the relations between dynamical objects moving
in space-time. For a physicist every phenomenon should be
described in terms of observable quantities and physical processes,
that is in an operational manner. From this point of view it is
clear that, without matter fields to probe the spacetime geometry, it is very hard to really `observe' our
quantum spacetime and  address properly the `semiclassical' issues
raised previously.

In order to promote these statements into physics,
we need to propose explicit quantum gravity observables when
matter fields are present. Remarkably, such observables are very
easy to construct and have been right in front of our eyes for a long
time: they are simply the Feynman diagrams. In order to be precise,
let us define what we mean.
A closed\footnote{We restrict ourselves in this paper to closed
Feynman diagrams. } Feynman diagram is a purely combinatorial data,
namely an abstract graph whose edges are colored by
representations of the Poincar\'e group (mass and spin), ends of
edges are labeled by representations of the Lorentz group
contained in the Poincar\'e group representation, and vertices are
labeled by intertwiners of the Lorentz group \cite{Weinberg}. In
the simplest case (spin zero), we often look only at the subclass
of Feynman diagrams with all Lorentz representations equal to the
trivial one, and we don't talk about this label - this is enough if
we only want to describe perturbative expansion of non-derivative
interactions.

Given such data, denoted by $\Gamma$, and a space-time manifold equipped
with a metric $g$, we can compute the Feynman amplitude
$I_\Gamma(g)$. Of course, what we are interested in is a sum
of Feynman diagrams, for instance those of fixed valency and given
degree, as generated by a field theory at given order of
perturbation theory. In order to keep the exposition simple we
will refer to individual Feynman diagrams, keeping in mind this important remark.
The object of interest is the quantum gravity observable \be
\widetilde{I}_{\Gamma}(l_{p}) = \int {\cal D}g\,
e^{\frac{i}{l_{p}}S(g)}I_\Gamma(g), \ee where the integral is over
the space of metrics, $l_{p}$ is the Planck length and $S(g)$ is
the gravity action. Of course we have to be able to make sense of
this path integral, and spin foam models aim to give a
background independent way of computing this amplitude. No
definite proposal is available yet in this framework. However, even
in the absence of a definite proposal, we strongly claim that we
can still propose a falsifiability test of the spin foam hypothesis as a valid candidate for a theory of quantum gravity.

 The main point is that whatever the quantum gravity amplitude is, one should be able to recover
 from it usual field theory when quantum gravity effects are negligible; that is,
 $\widetilde{I}_{\Gamma}(l_{p}) $ should admit a perturbative expansion
 \be
 \widetilde{I}_{\Gamma}(l_{p}) =
 I_{\Gamma}^{(0)}+ l_{p}I_{\Gamma}^{(1)}+l_{p}^2I_{\Gamma}^{(2)}
 +o(l_p^{2}).
 \ee
 Moreover, the first term $ I_{\Gamma}^{(0)}$ in the expansion should be the evaluation of usual Feynman diagram in the gravity vacuum state,
 namely flat space - or de-Sitter or Anti-de-Sitter if a cosmological constant is included in the gravity action.
 This is  a mandatory constraint on any proposal for a background independent approach to quantum
 gravity if we want to make the link with experiments and the highly successful effective field theory point of view.

  This provides a non-trivial first step falsifiability test on the spin foam hypothesis.
  Indeed, this hypothesis implies\footnote{We refer the reader to \cite{review} for introduction and review
  on spin foam models.} that $\widetilde{I}_{\Gamma}(l_{p})$, and hence  $ I_{\Gamma}^{(0)}$, should be written as
  a combinatorial state sum model depending on the choice of a triangulation $\Delta_{\Gamma}$
  adapted to the Feynman graph, and of a coloring of the faces and edges of $\Gamma$ by Lorentz group
  representations; that is
\be
 \widetilde{I}_{\Gamma}(l_{p})= \sum_{j_{f},j_{e}}  \prod_{f}A_{f}(j_{f})  \prod_{e}A_{e}(j_{e}, j_{f})
 \prod_{v}A_{v}(j_{e}, j_{f}) O_\Gamma(j_f,j_e)
 \ee
 where $f, e, v$ denote faces, edges and vertices of the 2-complex $\mathcal{J}_{\Delta}$ dual to the triangulation, and
$ A_{f},A_{e},A_{v}$ are local face, edge and vertex
amplitudes depending on the spins which are summed and represent quantum gravity fluctuations.
$O_\Gamma$ is an observable characterizing the coupling of Feynman diagram from insertion of matter.
 To any field theorist familiar with Feynman graphs, this seems to be a structure quite removed from
 anything a Feynman integral looks like.

The second consequence of the spin foam hypothesis follows from
the work \cite{FA} where a new background independent approach to
quantum gravity perturbation theory was proposed in the language
of spin foam model. In this approach, the starting point is to
write 4d gravity as a perturbation of a topological $BF$ theory
based on the de-Sitter group for positive cosmological constant.
The perturbation parameter $G_{N}\Lambda$ is dimensionless and the
perturbation theory transmutes gauge degree of freedom into
physical degrees of freedom in a controlled way, order by order.
In particular this means that the theory becomes topological in the limit $G_{N}\to
0$.
It has also been shown in this context that the coupling to matter particles can be explicitly performed by
computing expectation value of Wilson lines observables \cite{FAJ}, which are the most natural gauge invariant observables in this formulation.

 The main consequence of interest to us from these works is the fact that, not only Feynman diagram
amplitudes should be written as expectation values of certain natural observables in a spin foam model, but, moreover, the
corresponding model should be a topological spin foam model based on a Poincar\'e $BF$ theory.

So in summary, the spin foam hypothesis implies that usual Feynman
graph can be expressed as the expectation value of certain
observables in a topological spin foam model based on the Poincar\'e
group. The validity of such a statement is for us a non-trivial
check in  support of the spin foam hypothesis. The check is fourfold:
first, spin foam should  arise naturally in Feynman integrals; second,
the spin model should agree with the structure predicted by
\cite{FA,FAJ};  third, it should confirm the idea that the limit
$G_{N}\to 0$ is a limit where gravity becomes topological; and fourth the Feynman diagram observables
should be understood as a Wilson lines (or more generally spin networks) expectation value in this spin foam model.

\medskip

In this paper, we show that the first three conditions are indeed satisfied. We will take a very conservative approach not
relying on any hypothesis about quantum gravity dynamics. Instead, we will carefully
study the structure of Feynman integrals, and show that they can indeed be written as the expectation value of certain observables
in an explicit topological spin foam model.
The idea of our derivation is to  consistently erase the information about flat space geometry from the Feynman integral
and encode this information in terms of a choice of quantum amplitudes that should be summed over, and which {\it dynamically} determine
flat space geometry. In doing so, a triangulation, and a specific spin foam model living on it, are naturally found;
this allows us to express usual field theory amplitude in a background independent manner.
The idea that spin foam models code, in a background independent manner, the integration measure
viewed by Feynman diagrams was formulated for the first time in \cite{BarrettFD} and \cite{PRIII} in the context of 3d-gravity.

An analysis similar to the one done here has already been performed in 3d \cite{hqg1},
where it has been shown that the corresponding spin foam model
is constructed in terms of 6j symbols of the 3d Euclidean group
for flat space.
%( or of a deformation of the Euclidean group for non zero cosmological constant).
The deformation of this spin foam model using quantum group naturally leads to a formulation
of Feynman diagram coupled to 3d quantum gravity amplitudes \cite{PRIII,PRIII1}.
This corresponds to a deformation of field theory carrying a deformed action of the Poincar\'e group.

The strategy followed here is, to some extent, analogous to the one followed by Polyakov \cite{Polyakov},
when he showed that Feynman diagrams can be rewritten as worldline integrals.
Such an interpretation leads to powerful partial resummation of Feynman diagrams \cite{Bern}.
Moreover, it leads to a natural proposal for deforming the structure of quantum field theory in terms of a dimensionfull parameter,
that is to consider worldsheet (instead of worldline) integrals - hence string theory. In our case we have written 4d Feynman diagram in terms of a specific spin foam model,
and this hopefully opens a new way to think about consistent dimensionfull deformations of field theory structure. From the field theory perspective, such deformations due to the coupling of matter fields to gravity, might serve as natural regulator of the infrared and ultraviolet singularities that faces quantum field theory on fixed background metric.

This reformulation of Feynman graph amplitudes does not a priori simplify the computation of usual Feynman diagrams,
but, since it is based on a topological field theory,
it allows to give a natural generalization of the definition of Feynman amplitudes in the context where the underlying manifold admits
a non-trivial topology.

\medskip

The paper is organized as follows. In section \ref{1}, we show how, from a reformulation of Feynman integrals in terms of relatives distances between
vertices, the background geometry can be induced dynamically in the amplitudes.
This analysis will allow us to formulate our main statement, namely that Feynman amplitudes are the evaluation of observables
for an explicit spin foam model. The section \ref{symmetries} focuses on a
careful study of this model: the complete identification of its symmetries and the expression of the gauge-fixed partition function.
The proof of our statement is then given in section \ref{proof}: first the gauge-fixed model is shown to be invariant under Pachner moves; then
for any graph $\Gamma$, the so-called Feynman graph observable is defined,
which breaks part of the symmetry of the model, promoting gauge degrees of freedom into dynamical degrees of freedom; finally
the expectation value of this observable is shown to reproduce the Feynman amplitude associated the graph $\Gamma$.
Eventually, in section \ref{alg}, the algebraic
and physical interpretation of the Feynman graph spin foam model is discussed.

%%%%%%%%%%%%%%%%%%%%%%%%%%%%%%%%%%%%%%%%%%%%%%%%%%%%%%%%%%%%%%%
\section{Dynamical geometry in Feynman amplitudes} \label{1}
%%%%%%%%%%%%%%%%%%%%%%%%%%%%%%%%%%%%%%%%%%%%%%%%%%%%%%%%%%%%%%%

We restrict our study to the case of closed Feynman diagrams that arise in the context of QFT in flat Euclidean space-time.
A  Feynman amplitude for a scalar field takes the form
\beq \label{feyamp}
I_{\Gamma} = \int_{\RR^4} \extd^4 x_1 \cdots \extd^4x_n\, \OO(|\vec{x}_i - \vec{x}_j|), \quad \quad
\OO = \prod_{(ij)\in \Gamma}G^F(\vec{x}_i - \vec{x}_j)
\eeq
$\Gamma$ is the Feynman graph, $\vec{x}_i, \, i=1, \cdots n$ denotes the positions in $\RR^4$ of the $n$ vertices of the graph.
The product is over all edges of $\Gamma$ and  $G^F$ is the Feynman propagator\footnote{We will work with a regulated form of the Feynman propagator
in order to avoid divergences. It is also understood that the volume of $\RR^{4}$ is divided out from this integral: this is achieved by
not integrating over one of the $x_{i}$. }. In this section we show how to write this amplitude
as a sum over labels living on a specific triangulation of a $4$-dimensional ball, of the product of propagators. This result is established
in every dimension in our previous work \cite{hqg1}, and here we summarize the main arguments leading to it. We then describe, through a simple example,
how flat geometry can be dynamically implemented in (\ref{feyamp}).

\medskip

The usual way to express QFT amplitudes involves Lebesgue measures in $\RR^4$, which explicitly carry information about flat geometry.
The product of propagators depending only on the distances between vertices of the graph, the integrand in (\ref{feyamp}) is invariant under the action
of the Euclidean   group $ISO(4) = SO(4) \times \RR^4$. A first idea here is to gauge out this symmetry and express the integral in terms of the
invariant measure, acting on the space of functions of the relatives distances $l_{ij} = |\vec{x}_i - \vec{x}_j|$. The invariant measure can be
constructed out of the Lebesgue measure as follows. Consider first the case of four points, which are the vertices of
a tetrahedron in $\RR^4$. The relative position of these points is fully specified by the edge lengths of the tetrahedron.
The Lebesgue measure splits then into the Haar measure $\extd^4 a \extd{\Lambda}$ of the Poincar\'e group and a product of $l_{ij} \extd l_{ij}$:
\be \label{meas4}
\extd^4x_1\cdots \extd^4x_4 = \extd^4 a \extd{\Lambda} \prod_{i<j}l_{ij} \extd l_{ij},
\ee
With an additional point $x_5$, one can form a $4$-simplex $\sigma$, and write the Lebesgue measure in terms of the four edge lengths $l_{i5}$ as
\be \label{invTet}
\extd^4x_{5} = \sum_{\epsilon} \frac{\prod_{i=1}^4 l_{i5}\extd l_{i5}}{\mathcal{V}(l_{ij})}
\ee
In this formula $\VV$ is the volume\footnote{For a flat D-simplex $\sigma$ in $\RR^D$, with vertices $\vec{x}_i$, $\VV_{\sigma}$
denotes the square root of the determinant $\mbox{det}(\vec{l}_i \cdot \vec{l}_j)$,
with $\vec{l}_{i} = \vec{x}_i - \vec{x}_1, \, i = 2 \cdots D+1$, and it therefore equals $D!$ times the volume of the simplex. Abusing terminology
the quantities $\VV$ will be called `volume' in all the paper. For the particular case $D=2$, the simplex is a triangle $F$, the volume is an area
and the notation $\A_F$ is
used instead of $\VV$.} of the simplex, and $\epsilon \in \{\pm1\}$
labels its orientation. The simplex constitutes
the simplest triangulation $\Delta_1$ of a $4$-ball, without any internal face. For general values of $n=4+k$,
the invariant measure is obtained recursively: if $x_1, \cdots, x_{4+p}$ are the vertices of a triangulation $\Delta_p$ of a $4$-ball,
without any internal face,
on can choose a tetrahedral face of $\Delta_p$ and connect an additional point $x_{4+(p+1)}$ to its four vertices, to form a new triangulation $\Delta_{p+1}$,
and compute the Lebesgue measure by using (\ref{invTet}). Eventually, the measure
is expressed in terms of the edge lengths and orientations of the $4$-simplices of the triangulation $\Delta_k$.
\be
   \mathrm{d}^4x_1\cdots\mathrm{d}^4x_{4+k} = \extd^{4}a \extd\Lambda \sum_{\epsilon \in \{\pm 1\}^k} \prod_{e\in \Delta_{k}}l_e\extd l_e
   \prod_{\sigma\in\Delta_{k}}\frac{1}{\mathcal{V}_{\sigma}}
\ee
With this formula we can express the Feynman integrals purely in term of edge lengths of $\Delta_{k}$,
up to an overall $SO(4)$ volume factor that we drop out from now on.

$\Delta_k$ is a triangulation of a $4$-ball, with $4+k$ vertices, such that all vertices, edges and faces lie on the boundary. Let us emphasize that
any triangulation of this kind can be built recursively by the procedure described above, once an ordering of the vertices is chosen.
This can be seen by noticing that such triangulations are those for which the dual $1$-skeleton is a connected
$5$-valent tree (containing no loops), with open ends.
Now in order to draw such a tree, one first chooses one of its vertex $v_0$, called the `root' of the tree, and draws the four edges meeting at $v_0$;
one of these edges connects $v_0$  to a second vertex $v_1$, and one draws the three additional edges meeting at $v_1$, and so on. By using the duality
between tree and triangulation (namely, vertices of the tree are dual to $4$-simplices, and edges are dual to tetrahedra)
we see that the building procedures of trees on one hand, and triangulations
$\Delta_p$ on the other, are identical.

Given such an abstract triangulation $\Delta_k$, the data $\{l_e, \epsilon_{\sigma}\}$ of edge and simplex labels defines a flat geometry on the triangulation,
that is, specifies the relative position of the vertices  in $\RR^4$. Consequently, distances between vertices which are not connected
by any edge of $\Delta_k$ are well defined functions of the labels. We denote symbolically $l_{e'}^{\epsilon}$ these functions;
the `prime' means that the edge $e'$ does not belong to the triangulation. The Feynman amplitude is finally given in terms of the invariant measure by
\beq
\label{Feyamp}
I_{\Gamma} = \int \prod_{e\in\Delta_{k}}l_e\extd l_e \sum_{\epsilon \in{\{\pm 1\}}^{k}}
\prod_{\sigma\in\Delta_{k}} \frac{1}{\mathcal{V}_{\sigma}} \OO(l_e,l_{e'}^\epsilon({l_e}))
\eeq
where the product of propagators depend both on the edge labels and the `missing' distances $l_{e'}$. In this expression the overall factor
corresponding to the gauge volume has been dropped.

\begin{figure}[t]
\includegraphics[width=6cm]{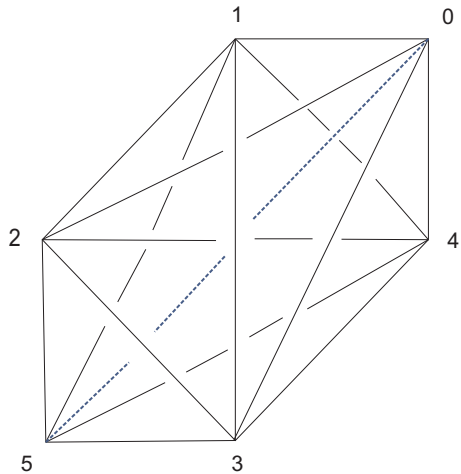}
\caption{A $5$-simplex defines a complex of two $4$-simplices $\sigma_0, \sigma_5$ sharing a tetrahedral face $\left[1234\right]$, as well as a complex of
four $4$-simplices $\sigma_1, \sigma_2, \sigma_3, \sigma_4$ sharing an edge $(05)$.}
\label{hexagone}
\end{figure}

\medskip
This formula is not enough since there is still an explicit flat geometry dependence encoded in the functions $l_{e'}$.
In order to go further, lets consider the following example where the Feynman graph $\Gamma$ forms the $1$-skeleton of a $5$-simplex,
with $6$ vertices and $10$ edges connecting all these vertices together, as shown in Fig.\ref{hexagone}.
We denote by $\sigma_j$ the $4$-simplex obtained by dropping the point $j$, by $\VV_j$ its volume and
$\epsilon_j$ its orientation.
The two $4$-simplices $\sigma_0, \sigma_5$, sharing a tetrahedron,
triangulate a $4$-ball, in such a way that all faces belong to the boundary, and that all vertices are connected to each other except for $0$ and $5$.
The product of propagators depends on the edge lengths $l_{ij}, (ij) \not= (05)$ of the triangulation,
as well as the distance between the points $0$ and $5$ which, as emphasized above, is a function of lengths $l_{ij}$ and orientations
$\epsilon_0, \epsilon_5$.

 For that particular case one can specify the dependence on the orientations. Indeed, it is possible to choose
conventions\footnote{For more details we refer the reader to the appendix of  \cite{hqg1}.} on orientations such that,
 when   two $4$-simplices
embedded in $\RR^4$ and
sharing a tetrahedron $\tau$ have identical orientations, then the points opposite to the common tetrahedron in each $4$-simplex
do not belong to the same half-space defined by the hyperplane spanned by $\tau$.
It is easy to convince oneself that if all $l_{ij},\,(ij) \not= (05)$ are fixed, then $l_{05}$ can take two values
$l_{05}^{\pm}(l_{ij})$,
with $l_{05}^- < l_{05}^+$ and that the sign `$\pm$' coincides with the product $\epsilon_0\epsilon_5$.

According to (\ref{Feyamp}), the Feynman amplitude for this graph reads
\beq \label{ex}
I_{\Gamma} = \int \prod_{(ij) \not= (05)} l_{ij} \extd l_{ij} \sum_{\epsilon_0, \epsilon_5}
\frac{1}{\VV_0\VV_5} \OO(l_{ij},l_{05}^{\epsilon_0\epsilon_5}({l_{ij}}))
\eeq
The form of the function $l_{05}^{\epsilon}$ encodes the flatness of the geometry. Our goal is to show that this function can be
replaced by a free label $l_{05}$, and that the flat geometry can be induced dynamically. The keystone of the proof is the remarkable identity of measures
\beq \label{id4d}
\sum_{\epsilon_0,\,\epsilon_5} \frac{\delta(l_{05} - l_{05}^{\epsilon_0\epsilon_5})}{\mathcal{V}_0\mathcal{V}_5}
= \sum_{\epsilon_1,\,\epsilon_2,\,\epsilon_3} l_{05} \frac{\mathcal{A}_{045}}{\mathcal{V}_1\mathcal{V}_2\mathcal{V}_3}\,\delta(\omega^{\epsilon}_{045})
\eeq
where $\A_{045}$ is (2 times) the area of the triangle $\left[045\right]$.
In this identity all lengths $l_{ij}, \,i,j=1\cdots4$ are fixed, the length $l_{05}$ being free to fluctuate; the measures can be used to integrate  functions
$f(l_{05})$ of this label. The delta function in the right hand side is the $2\pi$-periodic delta function; its argument is the deficit angle of the face $045$
$$
\omega^{\epsilon}_{045} = \sum_{i=1}^3 \epsilon_i \theta^i_{045}
$$
where $\theta^i_{045}$ is the dihedral angles of the face $\left[045\right]$ in the $4$-simplex $\sigma_i$. This deficit angle, considered as a function
of $l_{05}$ (and the orientations), is the curvature, in the sense of Regge calculus, carried by the face. It vanishes modulo $2\pi$ if and only if
the complex of four simplices $\sigma_1, \cdots,\sigma_4$ can be mapped in $\RR^4$, in such a way to give the orientation $\epsilon_j$ to the simplex
$\sigma_j$.
By symmetry of the role of the four points $1\cdots4$, similar identities hold for any permutation of $(1234)$.
(\ref{id4d}) is the four
dimensional analogue of the identity used in \cite{hqg1}
Also the proof is similar, namely consists in identifying,
thanks to the formula (\ref{4d1}), (\ref{4d2}) ,
the two sides of the equalities to the functional
$4l_{05}\delta(G)$, where $G \equiv \VV^2$ is the square of the volume of the $5$-simplex $\left[0\cdots5\right]$.
%Such an identity also appears in the work of Korepanov \cite{Korepanov}.

Plugging (\ref{id4d}) into the Feynman integral (\ref{ex}) promotes the lengths $l_{05}^{\e_0\e_5}$ to a free label; the price to pay is a constraint
which plays the role of a projector on the space of flat geometries.
If one expands the delta function as a sum $2 \pi\delta(\omega)=\sum_{s \in \ZZ} e^{\imath s \omega}$, the Feynman amplitude takes the
following form:
\beq \label{new}
I_{\Gamma} = \frac{1}{2\pi} \int \prod_{(IJ)} l_{IJ} \extd l_{IJ} \sum_{s \in \mathbb{Z}} \A_{045}
\sum_{\epsilon_1, \epsilon_2, \epsilon_3} \frac{e^{\imath s \omega^{\epsilon}_{045}}}{\VV_1\VV_2\VV_3} \OO(l_{IJ})
\eeq
The integral is over all edges connecting the vertices $0,\cdots,5$. It represents therefore
a sum over all (not necessarily flat)  geometries of the simplicial complex.
\medskip

All  these considerations serve as a motivation for our main statement, which will be established later:
the Feynman amplitude (\ref{Feyamp}) can be written as the expectation value of an observable
\beq \label{Exp}
I_{\Gamma} = \langle \OO(l_e)\rangle_{\Delta}, \qquad \OO = \prod_{e \in \Gamma} G^F(l_e)
\eeq
for the spin foam model:
\beq \label{4dmod}
Z_{\Delta}= \frac{1}{(2\pi)^{|F|}} \int\prod_{e\in\Delta} l_e\mathrm{d}l_e \prod_{F\in\Delta}\mathcal{A}_{F}
\sum_{\{s_F,\epsilon_{\sigma}\}} \left( \prod_{\sigma}\, \frac{e^{\imath \epsilon_{\sigma} S_{\sigma}(s_F, l_e)}}{\VV_{\s}}\right)
\eeq
$\Delta$ is a  triangulation of a closed 4D-manifold  and $\Gamma$ is embedded into the one-skeleton of $\Delta$.
Edges are labelled by positive numbers $l_e$, summed over a domain where triangular inequalities are satisfied; faces are labelled by integers $s_F$.
We denote by $\epsilon_{\sigma}=\pm 1$ the orientation
of the simplex $\sigma$. The measure involves a product of area $\mathcal{A}_{F}(l_e)$ of all faces $F$, while
the action for each simplex is similar to the 4d Regge action \cite{Re}.
\be \label{action}
S_{\sigma}(s_{F},l_{e})= \sum_{F\in \sigma} s_{F} \theta_{F}^\sigma(l_e),
\ee
where $\theta_{F}^\sigma(l_{e})$ is the interior dihedral angle of the face $F$ in $\sigma$. $|F|$ denotes the number of faces of the triangulation.

As we will see,
the evaluation (\ref{Exp}) is independent of the choice of the triangulation  which contains $\Gamma$ as a subgraph.
Also, the model (\ref{4dmod}) is a state-sum version of the 4-manifold
invariant constructed  by Korepanov in the remarkable work \cite{KorI, KorII, KorIII}.
The observable $\OO$ is
the product of propagators in which the distances are replaced by the labels $l_e$ living on the graph $\Gamma$.
The equality (\ref{Exp}) is obtained if one restricts to trivial topologies, that is, if $\Delta$ triangulates the $4$-sphere
$\SSS^4$.

\medskip

Notice that, although the structure of the integrand in (\ref{new}) is similar to that of the model (\ref{4dmod}), they are not identical. In particular,
a product of the area of all the faces, a product of the volume of all the simplices and a sum of the labels of all the faces are taken over in the
latter, while, in the former,
only the area of the face $\left[045\right]$ appears, only the label of this same face is summed over, and the volume $\VV_{4}$ is missing.
The reason of such differences is the following: the integral (\ref{new}) has to be interpreted as expectation value for the \textit{gauge-fixed} model,
and therefore hides a gauge-fixing term and a Faddeev-Popov determinant. Hence, what we learn from the study of the example is actually twofold:
not only it leads to the explicit proposal (\ref{4dmod}) for our statement, but it also indicates that, in order to prove this statement,
we definitely need to identify the symmetries of the model and find the gauge-fixing prescriptions.

In the following, as suggested by the previous remark,
we will start by dwelling further into the meaning of the integration measure which appears in (\ref{4dmod}).
As is customary for theories with gauge symmetries, this measure should be understood as the naive measure modulo
gauge transformation, and defined using a Faddeev-Popov gauge-fixing procedure: we will study in detail the symmetries of our model
and construct explicitly the gauge-fixed measure. This analysis will allow us to define unambiguously the model.
We then prove the statement (\ref{Exp}) in the subsequent section.

%%%%%%%%%%%%%%%%%%%%%%%%%%%%%%%%%%%%%%%%%%%%%%%%%%%%%%
\section{Symmetries and gauge fixing}
%%%%%%%%%%%%%%%%%%%%%%%%%%%%%%%%%%%%%%%%%%%%%%%%%%%%%%

\label{symmetries}

The total action of the model reads
\beq \label{act}
S_{\Delta}[s_F,l_e]= \sum_{\sigma} \epsilon_{\sigma} S_{\sigma}(s_F, l_e) = \sum_{F} s_F \omega^\epsilon_F(l_e)
\eeq
involving the deficit angle of each face of the triangulation. In ways similar to the $3$d case \cite{hqg1}, symmetries mapping
classical solutions to classical solutions induce divergences in (\ref{4dmod}). Taking the view that these symmetries are gauge symmetries,
we want to write down explicitly the gauge-fixed model, free of these naive divergences.
The face labels $s_F$ are treated here as continuous variables.

%%%%%%%%%%%%%%%%%%%%%%%%%%%%%
\subsection{Classical solutions and zero modes} \label{symsol}
%%%%%%%%%%%%%%%%%%%%%%%%%%%%%

The equations of motions  corresponding to the action (\ref{act}) are given by
\beqa
0 &=& \frac{\delta S}{\delta s_F} = \omega^{\epsilon}_F(l) \quad \forall \, F \\
0 &=& \frac{\delta S}{\delta l_e} = \sum_{\sigma} \epsilon_{\sigma} \left( \sum_{F \subset \sigma} s_F \frac{\delta \theta^{\sigma}_F}{\delta l_e}\right)
 = \sum_F s_F \frac{\delta \omega^{\epsilon}_F}{\delta l_e} \quad \forall \,e
\eeqa
The first equation expresses the flatness condition. A set of labels $\{l_e^o\}$ is solution if the simplicial
complex $\Delta$ can be locally mapped in $\RR^4$.
Solutions of the second equation are provided by the Schl\"afli identity for the flat $4$-simplex, which implies:
\beq
\sum_{F \subset \sigma} \A_{F} \frac{\delta\theta^{\sigma}_F}{\delta l_e} = 0
\eeq
Thus, we see that $s_F^o = \alpha \A_F(l_e)$ are solutions of the equations of motion, $\alpha$ being an arbitrary constant.
 Remarkably, by inserting
this solution into (\ref{act}) one recovers the Regge action of discrete $4$d gravity
$$ \mathcal{S}_R = \alpha \sum_F \A_F \omega^{\epsilon}_F(l_e). $$
In the following a solution $(l_e^o, s_F^o = \A_F(l_e^o))$ is called a Regge solution.

We want to study the zero modes of the model, namely infinitesimal deformations $\delta l_e, \delta s_F$ of the labels which belong to the kernel
of the Hessian $\delta^2 S$, computed on shell. The system of equations which characterize this kernel is
\beqa \label{h1}
\sum_e \frac{\delta\omega_F}{\delta l_e} \delta l_e &=& 0 \quad \forall \, F \\ \label{h2}
\sum_F \frac{\delta\omega_F}{\delta l_{e'}} \delta s_F + \sum_{F, e} s_F \frac{\delta^2 \omega_F}{\delta l_e \delta l_{e'}} \delta l_{e}
&=& 0 \quad \forall \, e'
\eeqa
where the label $\epsilon$ has been dropped for clarity. One can reorganize the left hand side of the second equation by using derivatives of the
Schl\"afli identity, which
yields
\beq
0 = \sum_{\sigma} \epsilon_{\sigma} \frac{\delta}{\delta l_{e'}} \left[\sum_{F \subset \sigma} \A_{F} \frac{\delta\theta^{\sigma}_F}{\delta l_e} \right]
 = \sum_F \frac{\delta \A_F}{\delta l_{e'}} \frac{\delta \omega_F}{ \delta l_{e}} + \sum_F \A_F \frac{\delta^2 \omega_F}{ \delta l_e \delta l_{e'}}
\eeq
Equation (\ref{h2}) can then be written as
\beq
\sum_F \frac{\delta\omega_F}{\delta l_{e'}} \delta (s_F - \A_F)  + \sum_{F, e} (s_F - \A_F)  \frac{\delta^2 \omega_F}{\delta l_e \delta l_{e'}} \delta l_{e}
= 0
\eeq
where the symbol $\delta \A_F$ denotes the variation $\sum_e \frac{\delta \A_F}{\delta l_e} \delta l_e$ of the area. Now if one restricts to
fluctuations around Regge solutions, one obtains eventually the equations satisfied by the zeros modes
\beqa \label{H1}
\sum_e \frac{\delta\omega_F}{\delta l_e} \delta l_e &=& 0 \quad \forall \, F \\ \label{H2}
\sum_F \frac{\delta\omega_F}{\delta l_{e}} \delta s_F &=& 0 \quad  \forall \, e
\eeqa

These equations characterize two independent symmetries $l_e \to l_e + \delta l_e$ and $s_F \to s_F + \delta s_F$ which we now identify and then gauge-fix.

%%%%%%%%%%%%%%%%%%%%%%%%%%%
\subsection{Edge symmetry}
%%%%%%%%%%%%%%%%%%%%%%%%%%%
\label{SymmVert}

The symmetry of edge labels $l_e$ is similar to the one arising in $3d$ and has a simple geometrical interpretation.
 Starting from
a stationary point of the action, the variations $\delta l_e$ satisfying (\ref{H1}) are those for which the deficit angles remain
invariant. In other words, given a set of labels $\{l_e\}$ that define a flat geometry, solutions of (\ref{H1}) are such that the set of labels
$\{l_e + \delta l_e\}$ also define a flat geometry.
Since the geometry of the starting configuration is flat, one can embed the neighborhood of each vertex in $\RR^4$.
Variations of the labels that do not modify the geometry arise from infinitesimal moves of the vertices in $\RR^4$ - and therefore they are generated by
4-vectors $\vec{\alpha}_v$ attached to each vertex of the triangulation.

\begin{figure}[h]
\begin{center}
\includegraphics[width=6cm]{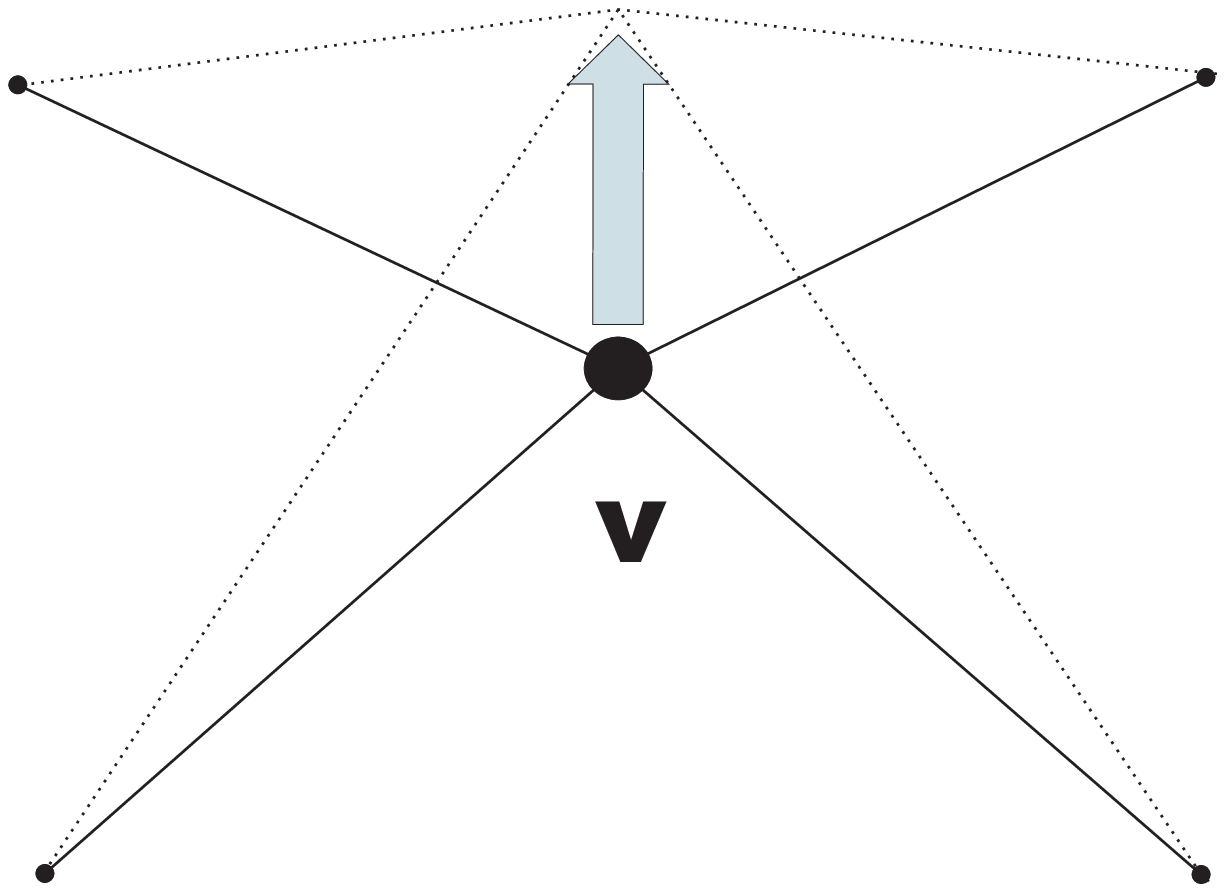}
\label{vmove}
\end{center}
\end{figure}

Given a vertex $v$ and an
infinitesimal $4$-vector $\vec{\alpha}_v$ associated to it, the corresponding variations of the labels $l_e$ are defined to be $0$ for each edge $e$
that does not touch the vertex $v$, and, for each edge $e$ touching $v$, to be the projection of $\vec{\alpha}_v$ in the direction defined by $e$, namely
$$
\delta l_e(\vec{\alpha}_v, l_e) \equiv - \vec{\alpha}_v \cdot \frac{\vec{l}_e}{l_e}
$$
where $\vec{l}_e$ is the vector represented by the edge $e$ when $v$ is placed at the origin.

The gauge-fixing is performed by fixing the value of a subset of labels
and by taking into account the Faddeev-Popov determinants. Intuitively,
in order to eliminate the four components of the gauge parameter at each vertex $v$, we need to fix the labels $(l_j)$ of four edges sharing $v$. If one
chooses these four edges so that they belong to a same simplex $\sigma$ touching $v$, the
corresponding determinant can then be read out from the relation between the Lebesgue measure $\extd^4 \alpha_v$ and the variations $\delta l_j$
computed in section (\ref{invTet})
$$
\extd^4\vec{\alpha}_v = 2\frac{\prod_{j=1}^4l_j\extd l_j}{\VV_{\sigma}}
$$
where the factor $2$ is due to the summation over the values of the orientation for $\sigma$.

The  gauge-fixing procedure goes as follows.
 We first choose $5$ vertices that form a $4$-simplex $\sigma_0$ in $\Delta$, and
then assign to every other vertex $v$ of the triangulation a $4$-simplex $\sigma_v$ to which this vertex belongs. Each of these vertices
provides four edges $e_v^1, \cdots, e_v^4$ , namely the four edges of $\sigma_v$ that meet at $v$. We impose the assignment
$A_l = (\sigma_0, \{\sigma_v\}_{v \notin \sigma_0})$ to satisfy the admissibility condition that
$e_v^i \not= e_{v'}^j$ for every couple $(v, v')$ of distinct vertices that are not in $\sigma_0$;
this condition insures that no edge is picked up more than once in the procedure.
Such an assignment can be constructed recursively
by using a maximal tree in the 1-skeleton dual of the triangulation \cite{hqg1}.

Given an admissible assignment $A_l$, we say that a $4$-simplex $\sigma$ belongs to $A_l$ if either $\sigma = \sigma_0$ or $\sigma = \sigma_v$
for some vertex $v$ not in $\sigma_0$,
and that an edge $e$ belongs to $A_l$ if either $e \in \sigma_0$ or
$e \in \sigma_v$ and $e$ admits $v$ as one of its vertices.
The gauge-fixing terms and Faddeev-Popov determinants associated to the symmetry (\ref{H1}) read then
\beq \label{gfsym1}
\delta_{GF}^{A_l} = \prod_{e \in A_l} \delta(l_e - l_e^o), \qquad D_{FP}^{A_l} =
\frac{1}{2^{|v|-3}} \frac{\prod_{\sigma \in A_l} \VV_{\sigma}}{\prod_{e \in A_l} l_e}
\eeq
where $|v|$ is the number of vertices of $\Delta$ and $l_e^o$ arbitrary fixed values of the labels.

%%%%%%%%%%%%%%%%%%%%%%%%%%%
\subsection{Face symmetry}\label{facesym}
%%%%%%%%%%%%%%%%%%%%%%%%%%%

We now want to deal with the symmetry of the face labels $s_F$. We will identify generators as being 3-vectors $\vec{\beta}_e \in \RR^3$ attached
to the edges of the triangulation.

Variations $\delta s_F$ that satisfy (\ref{H2}) can be described as follows.
Let $e$ be an edge of the triangulation. For a given configuration of the $l$'s labels inducing a flat geometry, the complex of $4$-simplices sharing
$e$ is mapped in $\RR^4$ and one can consider the vector space $e^{\bot} \simeq \RR^3$ orthogonal to the straight line spanned by
the edge. Next, we attached to $e$ an infinitesimal $3$-vector $\vec{\beta_{e}} \in e^{\bot}$, and consider the \textit{translation} in
$\RR^4$ of the edge $e$ by $\vec{\beta}_e$, that is, the translation of all the points of $e$ by the same vector
$\vec{\beta}_e$.

\begin{figure}[h]
\includegraphics[width=5cm]{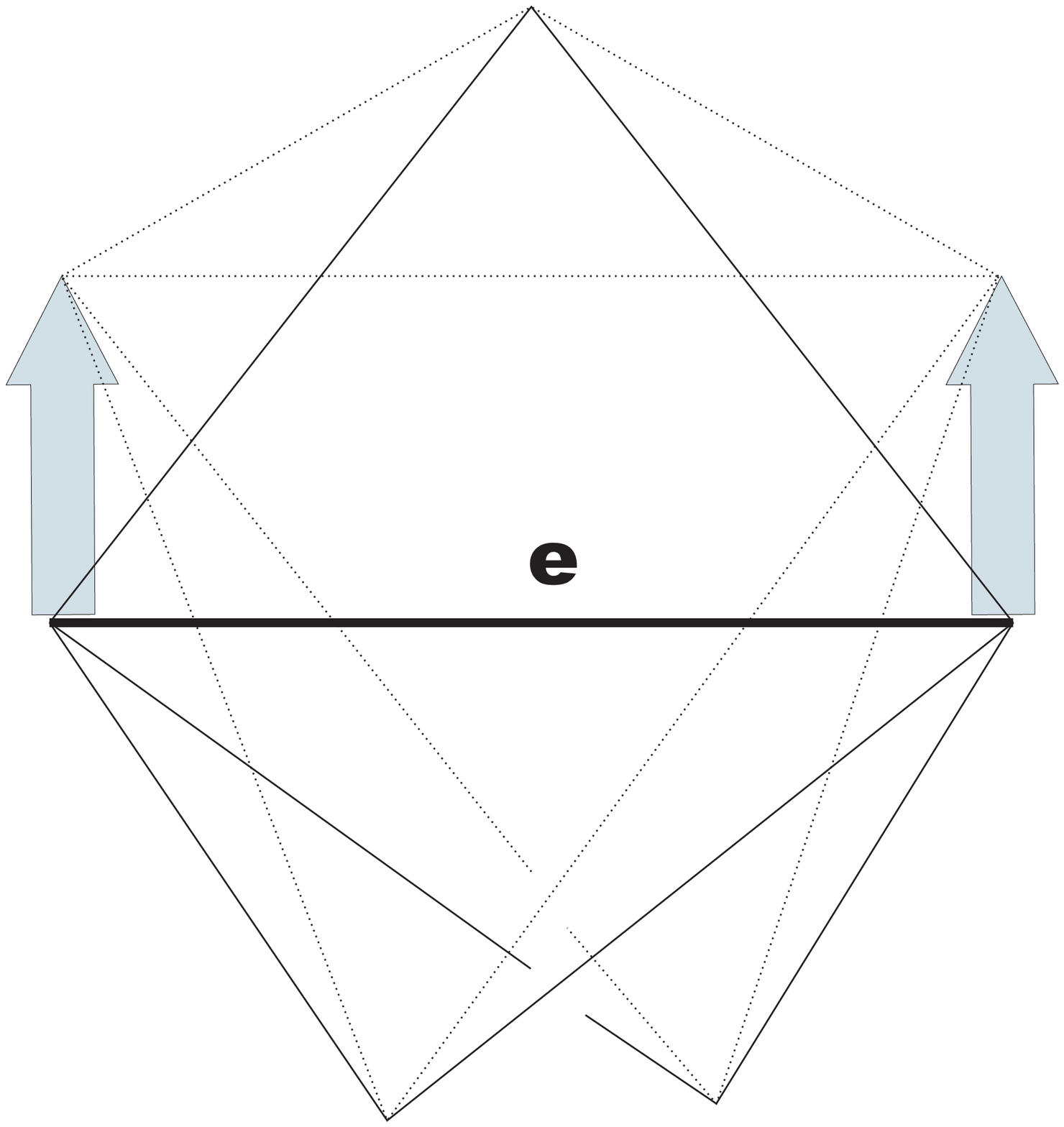}
\label{emove}
\end{figure}

This translation deforms each triangle $F$ to which $e$ belongs. In particular, it modifies
%the area $\A_F(l_e^o)$ of the triangle $F$; since the length $l_e$ is invariant under such a translation,
the height $h_F(l_e)$ associated to $e$ in $F$, that is, the height of the point opposite to the edge $e$ in $F$. The variation of this height induced
by the translation of $e$ reads
\beq \label{heights}
\delta h_F(\vec{\beta}_{e}, l_{e}) = - \vec{\beta}_e \cdot \frac{\vec{h}_F}{h_F}
\eeq
where $\vec{h}_F$ is the vector represented by the height when its intersection with $e$ is placed at the origin.
Then for any face $F$ of the triangulation, we define the transformation $s_F
\rightarrow s_F+\delta s_F$ generated by $\vec{\beta}_{e}$ as
\beq \label{transf}
\delta_e s_F= \begin{cases}
l_e^{-\frac23}\delta h_F(\vec{\beta}_e, l_e) \quad & \text{if $F \supset e$} \\
0 \quad & \text{if not}
\end{cases}
\eeq

\begin{figure}[t]
\includegraphics[width=10cm]{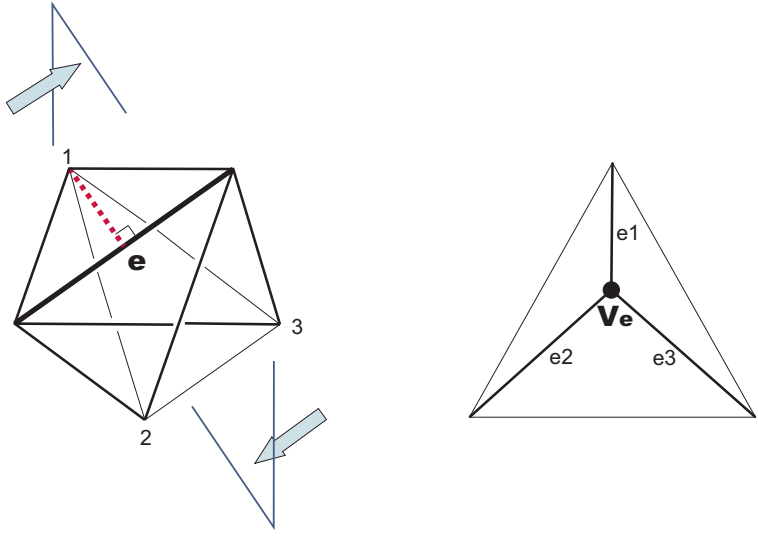}
\caption{The orthogonal projection of a simplex $\sigma \supset e$ onto the 3-space $e^{\bot}$, maps the edge $e$ to the vertex $v_e$ and
the three faces $F_i$ adjacent to $e$ to three edges $e_i$ meeting at $v_e$. The dashed line represents the height associated to $e$ within
the triangle $F_1$.}
\label{proj}
\end{figure}

In order to show that the variations $\delta_e s_F$ defined above\footnote{The unnatural prefactor $l_e^{-\frac23}$ is chosen in order to
simplify the computation of the determinant. Geometrically it would have made more sense to define $\delta_e s_F = l_e h_F$ since, on-shell, we identify
$s_F$ with an area. The key remark is that if $\delta_e s_F$ satisfies (\ref{H2}) then $\tilde{\delta}_e s_F \equiv f(l_e) \delta_e s_F$ satisfies
(\ref{H2}) as well. Accordingly, we are free to redefine the transformation of $s_F$ at our convenience, using any $f$.
The final result is independent of this choice.}
satisfy (\ref{H2}), let us mention a preliminary geometrical result.
Let $\sigma$ be one of the $4$-simplices to which the edge $e$ belongs.
The data of labels and orientation allows us to map $\sigma$ in $\RR^4$.
We consider the orthogonal projection $P: \RR^4\rightarrow\RR^3$ onto the space $e^{\bot}$.
P maps $e$ onto a vertex $v_e$, the three triangles $\{F_i,\,i=1,2,3\}$ meeting at $e$ onto
a triplet of edges $\{e_i\}$ meeting at $v_e$, and the simplex $\sigma$ itself onto a tetrahedron $\tau$ to which $e_i$ and $v_e$ belong, as shown in
Fig.\ref{proj}.
Note that the length of the edge $e_i$ equals the height $h_i$ associated to $e$ in the triangle $F_i$. Then the result is the following:
for all $i$,
the 4d dihedral angle $\theta^{\sigma}_{F_i}$ of the face $F_i$ in $\sigma$ equals the 3d dihedral angle $\theta^{\tau}_{e_i}$ of the edge $e_i$ in $\tau$.
This correspondence is geometrically clear:
the dihedral angle of $F_{i}$ is $\pi$ minus the angle between  the normal vectors to the two tetrahedra meeting at $F_{i}$.
They are orthogonal to the  edge $e$, thus the dihedral angle is given by the angle between their orthogonal projection.
These normal vectors project onto normals of the faces meeting at $v$, so that their angle is $\pi$ minus
the dihedral angle of the projected tetrahedron.

As we have seen, the transformation generated by a vector $\vec{\beta}_e$ attached to the edge $e$ affects only
the labels $s_F$ of the faces $F$ to which $e$ belongs.
Therefore, in order to show the equality (\ref{H2}), one can restrict the sum to the
faces sharing $e$. Thus, we need to show
\beq
\sum_{F\supset e} \frac{\delta\omega_F}{\delta l_{e'}} \delta s_F = 0 \quad  \forall \, e'.
\eeq
Let $F$ be a face such that $F \supset e$.
We consider the collection $\mathcal{C}_{F} = \{\sigma^j\}$ of all $4$-simplices around $F$:  $F\subset\sigma^j$.
For each $j$, $\sigma^j$ is embedded in a copy of $\RR^4$ and one can define, as before, the projection $P^j$ onto the sub-space
$e^{\bot j}$, the vertex
$v_{e}^j\equiv v_{e}$ image of the edge $e$, and the tetrahedron $\tau^j$ image of the simplex $\sigma^j$.
We also denote by  $e_F^j\equiv e_F$ the common image of $F$ by the projections $P^j$.
The collection $\{\sigma^j\}$ form a complex of $4$-simplices around the face $F$, and we see that the projections $P^j$ define
a complex $\{\tau^j\}$ of tetrahedra around the edge $e_F$.
Now thanks to the correspondence between 3d and 4d dihedral angles mentioned above,
the deficit angles $\omega_F$ of the face $F$ and $\omega_{e_F}$ of the edge $e_F$ equal each other:
\beq
\label{defangle}
\omega_F  \equiv  \sum_j \epsilon_j \,\theta_F^{\sigma_j} = \sum_j \epsilon_j \,\theta_{e_F}^{\tau_j} \equiv \omega_{e_F}
\eeq
The deficit angle $\omega_F$ is a function of the edge lengths $\{l_{e'}\}$ of the 4d complex $\mathcal{C}_F$, while the deficit angle $\omega_{e_F}$
depends on the labels $\{l_{e''}\}$ of the 3d complex $\mathcal{C}_{e_F}$ - $l_{e''}$ are well defined function of the $l_{e'}$:
$l_{e''} \equiv l_{e''}(l_{e'})$. Therefore, given an edge $e'$, differentiating (\ref{defangle}) with respect to the label $l_{e'}$ yields
\beq \label{der1}
\frac{\delta\omega_F}{\delta l_{e'}} = \sum_{e''} \frac{\delta\omega_{e_F}}{\delta l_{e''}} \frac{\delta l_{e''}}
{\delta l_{e'}}
\eeq
Then, let us first mention that the matrix
$\left(\frac{\delta \omega_e}{\delta l_{e'}}\right)_{e,e'}$ is symmetric, since\footnote{according to the Schl\"afli identity.}
it is the Hessian of the 3d Regge function $$\mathcal{S}_R^{(3)} = \sum_e l_e \omega_e.$$
Secondly, let us recall that the length $l_{e_F}$ of the edge $e_F$ equals the height $h_F$ associated to $e$ in the triangle $F$.
Consequently, multiplying (\ref{der1}) by the variation $\delta s_F = l_e^{-\frac23} \delta h_F$
and summing over $F \supset e$, lead to the following equalities, holding for every $e'$ of the triangulation:
\beqa
\sum_{F}\,\frac{\delta\omega_F}{\delta l_{e'}} \delta s_F
&=& \sum_{F, e''} \frac{\delta\omega_{e''}}{\delta h_F} \frac{\delta l_{e''}} {\delta l_{e'}} \delta s_F \\
&=&l_e^{-\frac23} \sum_{e''} \left( \sum_{F \supset e} \frac{\delta\omega_{e''}}{\delta h_F} \delta h_F \right) \frac{\delta l_{e''}} {\delta l_{e'}}
= l_e^{-\frac23}  \sum_{e''}\left( \sum_{e_F \supset v_e} \frac{\delta\omega_{e''}}{\delta l_{e_F}} \delta l_{e_F} \right) \frac{\delta l_{e''}} {\delta l_{e'}}
\eeqa
Now, on-shell, all the deficit angles vanish; all the spaces $e^{\bot j}$, as well as the projection $P^j$,
can be identified to each other $e^{\bot j} \equiv e^{\bot}$ and $P^j \equiv P$;
and the complex of tetrahedra $\tau_j$ can be mapped in $\RR^3$.
The term in parenthesis turns out to be the variation of the 3d deficit angle induced by a displacement, by a 3-vector
$\vec{\beta_e} \in e^{\bot} \simeq \RR^3$,
of the vertex
$v_{e}$; therefore it vanishes, and (\ref{H2}) is proved. Hence, we have identified the parameters $\vec{\beta}_e$
as generating the symmetry of the face labels.

\medskip

The gauge-fixing is performed by fixing a subset of the face labels while taking into account the Faddeev-Popov determinants.
By analogy with the edge symmetry, in order to eliminate the three components of the gauge parameter at each edge $e$,
we would need to fix the labels $(s_{F_i})$ of three faces sharing $e$. Lets consider a simplex $\sigma$ touching $e$
and the three faces $(F_i)$ of $\sigma$ that meet at $e$.
The `dimensional reduction' performed in the previous proof associates a tetrahedron $\tau$ to the simplex $\sigma$,
and three edges to the faces $F_i$; the lengths of these edges are also the heights $h_i$ associated to $e$ in $F_i$. The determinant
that corresponds to the fixing of the labels $(s_{F_i})$ can then be read out from the relation between the Lebesgue measure $\extd^3\beta_e$
and the variations $\delta h_i$
$$
 \extd^3 \vec{\beta}_{e} = \frac{\prod_{i=1}^3 h_i\extd h_i}{\VV_{\tau}}= \frac{\prod_{i=1}^3 \A_{F_{i}} \delta s_{F_{i}}}{\VV_{\sigma}}
$$
where $\VV_{\tau}$ is (3! times) the volume of the tetrahedron $\tau$, and where we made use, for the second equality, of
$h_{F}=l_{e},\, dh_{F}=l_{e}^{\frac{2}{3}} \delta s_{F}, \, \VV_{\sigma} = l_e \VV_{\tau}$, and  $\A_{F_i} = l_e h_{F_i}$.
 The determinant reads then
\beq \label{detAs2}
D^{\sigma}  = \frac{\VV_{\sigma}}{\prod_{i=1}^3\A_{F_i}}.
\eeq
 There is no factor $2$ in these expressions, since here we work with fixed
orientations for the simplices.

Still by analogy with the previous symmetry, we would want to assign,
to each edge $e$ of the triangulation, a simplex $\sigma_e$ to which $e$ belongs. Each of these simplices provides three faces $F_e^1, F_e^2, F_e^3$,
namely the three faces of $\sigma_e$ sharing $e$. Suppose that there exists an assignment $A_s = \{\sigma_e\}$ such that $F_e^i \not= F_e'^j$
for every couple
$(e, e')$ of distinct edges - this admissibility condition insures that no face is picked up more than once in the procedure.
Then the gauge-fixing terms and Faddeev-Popov determinant associated to the symmetry (\ref{H2}) are expected to be
\beq
\delta^{A_s}_{GF} = \prod_{F \in A_s} (2\pi) \delta_{s_F, s_F^o}, \qquad D_{FP}^{A_s} = \frac{\prod_{\sigma \in A_s} \VV_{\sigma}}{\prod_{F \in A_s} \A_F}
\eeq
where by definition $F \in A_s$ if $F \in \sigma_e$ and $F \supset e$ for some edge $e$.

A subtlety arises here however: it is in general impossible to choose an assignment $A_s$ satisfying
the  admissibility condition.
This can be seen by the following argument. Although this condition seems quite analogous to the admissibility condition
for the assignment $A_l$ associated to the edge symmetry, there is major difference between the comportment of these two conditions
under refinement of the triangulation, i.e under a $(1,5)$ Pachner move. Such a move consists in a subdivision of a simplex $\sigma_0$
into five $4$-simplices $\sigma_1, \cdots, \sigma_5$, providing one additional vertex, as well as five additional edges and ten additional faces.
Extending an admissible assignment $A_l$ requires to assign four edges to the new vertex, which can be picked up beyond the five new edges.
Extending an admissible assignment $A_s$, on the other hand, would require to assign to each of the five new edges a triplet of faces, without any
repetition of the faces.
 This means that  $3*5 = 15$ faces, adjacent to one of the new edges, are needed, whereas only ten are at our disposal.
 An over-counting of the faces, and, thus, of the labels that have to be fixed,
seems therefore to be unavoidable.
This problem, treated in the next section, is a reflection of the fact that the symmetry (\ref{H2}) generated by
$\vec{\beta}_{e}$   is a {\it reducible } symmetry \cite{BV}.
 It turns out, indeed, that the gauge parameters $\vec{\beta}_e$ are not independent,
which leads to an overestimation of the number of gauge degrees of freedom.

%%%%%%%%%%%%%%%%%%%%%%%%%%%%%%%%%%%
\subsection{Reducibility}
%%%%%%%%%%%%%%%%%%%%%%%%%%%%%%%%%%%

In this part we show that the action of the symmetry $s_F \to s_F +\delta s_F(\vec{\beta}_e)$ on the labels is not free, that is,
there exists non trivial transformations $\vec{\beta}_e \to \vec{\beta}_e +\delta \vec{\beta}_e$ of the gauge parameters, such that
\beq \label{secondstage}
\delta s_F (\vec{\beta}_e +\delta \vec{\beta}_e) = \delta s_F(\vec{\beta}_e).
\eeq
This symmetry of the gauge parameters is characterized infinitesimally by the following equations
\beq
\sum_e \frac{\delta s_F}{\delta \vec{\beta}_e} \delta\vec{\beta}_e = 0 \quad \forall \, F
\eeq
which provide dependence relations between the gauge parameters. As we will see, such transformations of the parameters $\vec{\beta}_e$
are generated by elements
$(\sigma_v)_{\mu\nu}$ of the Lie algebra $so(4)$ living on the vertices of the triangulation.
An analogous result can be found in Korepanov's work \cite{KorIII}.
Hence, the `true' degrees of freedom for the symmetry (\ref{H2}) are no longer
lists $\{\vec{\beta}_e\}_{e \in {\Delta}}$ of 3-vectors but rather \textit{orbits} of
such lists \textit{modulo} an action - which we will specify - of the Lie algebra elements.

\medskip

Transformations (\ref{secondstage}) can be described as follows. Let $v$ be a vertex of $\Delta$ and an element $\sigma_{\mu\nu} \in so(4)$
associated to it.
For a configuration $\{l_e\}$ of the $l$'s labels inducing a flat geometry, the complex of
$4$-simplices sharing $v$ can be mapped in $\RR^4$, in such a way that $v$ is placed at the origin.
The edges meeting at $v$ form then vectors $\vec{l}_e$ in $\RR^4$.
For any edge $e$ of the triangulation, we define the transformation
$\vec{\beta}_{e} \to \vec{\beta}_{e} + \delta \vec{\beta}_{e}$ generated by $\sigma_{\mu\nu}$ as
\beq \label{transf'}
\delta_v \vec{\beta}_{e} = \begin{cases}
l_e^{\frac23} \sigma\left(\frac{\vec{l}_e}{l_e}\right)  \quad & \text{if $e \supset v$} \\
0  \quad & \text{if not}
\end{cases}
\eeq
where $\sigma\left(\frac{\vec{l}_e}{l_e}\right) \equiv \sigma_{\mu \nu}\frac{l^{\nu}_{e}}{l_{e}}$ is the image of the unitary vector $\frac{\vec{l}_e}{l_e}$
by the operator which represents $\sigma_{\mu\nu}$ in $\RR^4$. Lets see why this transformation satisfies
(\ref{secondstage}).
Given a face $F$ to which the vertex $v$ belongs, let $e_a, e_b$ be the edges of $F$ meeting at $v$, and $l_a, l_b$ their lengths.
According to (\ref{transf}) and (\ref{transf'}), the variation $\delta s_F$ induced by the combined action of $\delta_v \vec{\beta}_{e_a}$
and $\delta_v \vec{\beta}_{e_b}$ can be written as
\beq \label{var1}
\delta s_F \equiv \delta_a s_F + \delta_b s_F =
- \frac{1}{l_a}\sigma(\vec{l}_a) \cdot\frac{\vec{h}^a_F}{h^a_F} - \frac{1}{l_b}\sigma(\vec{l}_b)\cdot \frac{\vec{h}^b_F}{h^b_F}
\eeq
where $h_F^a, h_F^b$ are the two heights  associated to $e_a$ and $e_b$ in $F$.
Note that it is always possible to find a scalar $\alpha$ such that $\vec{l}_{b}=\vec{h}_{a} +\alpha\vec{l}_{a}$
 and, since $\sigma$ is skew symmetric and $\A_{F}= l_{a}h_{a}= l_{b}h_{b}$, we have
 \be
 \delta_a s_F =\frac{\sigma(\vec{l}_{a})\cdot \vec{l}_{b}}{\A_{F}},\quad  \delta_b s_F =\frac{\sigma(\vec{l}_{b})\cdot \vec{l}_{a}}{\A_{F}},
 \quad  \delta_a s_F  + \delta_b s_F =0
 \ee
Hence, we see that the label $s_F$ transforms trivially under the action of $\delta_v \vec{\beta}(\sigma)$, which shows (\ref{secondstage}).

\medskip

The proper way to deal with the gauge fixing of the reducible symmetry is, first, to fix the symmetry of gauge parameters,
acting at the vertices
of the triangulation, and thus reduce the gauge degrees of freedom to an independent set of gauge parameters; and second, to
eliminate the remaining gauge degrees of freedom by the usual Faddeev-Popov procedure.
Notice that, by taking into account the reducibility of the symmetry,
the over-counting problem highlighted at the end of the last section no longer holds: under a refinement of the triangulation (move $(1, 5)$),
the action of an $so(4)$-element $\sigma_{\mu\nu}$ (6 components) attached to the new vertex has to be fixed beforehand; it is
expected to reduce the number of gauge parameters to
$15 - 6 = 9$. Then, the remaining gauge degrees of freedom are eliminated by fixing the labels of $9$ of the $10$ new faces.

In order to get an intuition of what the precise gauge-fixing prescriptions should be,
let us again consider a vertex $v$ of the triangulation and an element
$\sigma_{\mu\nu} \in so(4)$ associated to it. We suppose that the neighborhood of $v$ is mapped in $\RR^4$, the vertex being placed at the origin.
The element $\sigma_{\mu\nu}$ generates a transformation of the gauge parameters $\vec{\beta}_e$
living on the edges meeting at $v$. These edges define vectors $\vec{l}_e$ in $\RR^4$ and we see, with (\ref{transf'}),
that the variations of the gauge parameters is related to
the displacement $\delta \vec{l}_e \equiv \sigma (\vec{l}_e)$ of these vectors under the infinitesimal rotation $1_{4\times4} + \sigma_{\mu\nu}$. Now we know how to eliminate
rotational degrees of freedom that act on $4$-vectors: given four edges $\vec{l}_1, \cdots, \vec{l}_4$ meeting at $e$ and belonging to a simplex $\sigma_v$,
this elimination consists in fixing the direction of $\vec{l}_1$, restricting $\vec{l}_2$ to a fixed plane and $\vec{l}_3$ to a fixed hyperplane.
In other words,
the gauge-fixing of the action of $\sigma_{\mu\nu}$ can be performed by fixing the value of the three components of $\vec{\beta}_1$, two components
of $\vec{\beta}_2$ and one component of $\vec{\beta}_3$. If these six components are denoted by $\{\beta^k\}$, the determinant
$\Delta_2$ that results from the gauge-fixing of this `second-stage' symmetry,
is the determinant of a square matrix whose elements are derivatives of the $\beta^k$ with respect to the six independent components $\{\sigma^l\}$
of $\sigma_{\mu\nu}$:
$$
\Delta_2 = \mbox{det} \left[\frac{\pa \beta^k}{\pa \sigma^l}\right]
$$
Once the symmetry (\ref{transf'}) is fixed, one can use the six remaining components $\{\beta^j\}$
of the gauge parameters living on $\vec{l}_1, \cdots \vec{l}_4$ to fix the
labels $s_F$ of the six faces of $\sigma_v$ that share the vertex $v$. The determinant $\Delta_1$ that results from the gauge-fixing of this
`first-stage' symmetry, is the determinant of a square matrix whose elements are derivatives of the $s_F$ with respect to the $\beta^j$:
$$
\Delta_1 = \mbox{det}\left[\frac{\pa s_F}{\pa \beta^j}\right]
$$
The factors $\Delta_1$ and $\Delta_2$ are explicitly computed in Appendix \ref{DetComp} for a suitable
choice of gauge-fixing conditions.

This analysis shows how to treat the face symmetry acting at the edges of a simplex $\sigma_v$ that share a vertex $v$, while taking into account
its reducibility. Namely, this partial gauge-fixing is performed by fixing the variables $s_F$ labeling the faces of $\sigma_v$ that meet at $v$, and by
inserting of Faddeev-Popov determinant
$$
D^{\sigma_v} = \Delta_1 \Delta_2^{-1}.
$$
The form of this determinant is a typical feature of a reducible symmetry.
The first term $\Delta_1$ arises from the integration of usual
fermionic ghosts, while the second term $\Delta_2^{-1}$ arises from the integration of bosonic ghosts for ghosts \cite{BV}.
The product takes the  remarkably simple form, given by (\ref{Detfinal}):
\beq\label{Dv}
D^{\sigma_v} = \frac{\VV_{\sigma_v}}{\prod_F \A_F}
\eeq
where $\VV_{\sigma_v}$ is the volume of the simplex $\sigma_v$, and where the product is over the faces of the simplex which meet at $v$.

Let us suppose that the previous procedure is applied for every vertex $v$ of the triangulation.
That is for each vertex we choose a $4$-simplex $\sigma_{v}$ and fix the value of $s_{F}$ for all faces of $\sigma_{v}$.
The symmetry is therefore fully reduced,
the gauge parameters living on the edges which belongs to none of the $\sigma_v$ act now freely on the labels $s_F$ that are not fixed.
Given such an edge $e$, the symmetry generated by $\vec{\beta_e}$ is treated by fixing the variables of
three faces $F_e^1, F_e^2, F_e^3$
adjacent to $e$. As emphasized in  section (\ref{facesym}), if one chooses these three faces so that they belong to the same simplex $\sigma_e$,
the corresponding determinant reads
\beq\label{De}
D^{\sigma_e} = \frac{\VV_{\sigma_e}}{\A_{F_e^1} \A_{F_e^2}\A_{F_e^3}}
\eeq
Repeating this operation for every edge $e \notin \bigcup_{v\in \Delta} \sigma_v$ completes the full gauge-fixing of the symmetry (\ref{transf'}).

\medskip

The general gauge-fixing procedure of the reducible face symmetry is then as follows. We first define an admissible assignment
$A^{(1)}_{s} = (\sigma_0, \{\sigma_v\}_{v \notin \sigma_0})$ as in section \ref{SymmVert}, using a maximal tree in the 1-skeleton
dual to the triangulation \cite{hqg1}. A convenient choice is $A_s^{(1)} = A_l$.
As before we say that an edge $e$ belongs to $A^{(1)}_{s}$
if either $e \in \sigma_0$, or $e \in \sigma_v$ for some $v$ and $e$ admits $v$ as one of its vertices; likewise we say that a face $F$
belongs to $A^{(1)}_{s}$ if either $F \in \sigma_0$, or $F \in \sigma_v$ for some $v$ and $F$ admits $v$ as one of its vertices.
We then assign to every additional edge $e \notin A^{(1)}_{s}$ a $4$-simplex $\sigma_e$; it can be checked that, by construction,
 $\sigma_e \notin A^{(1)}_{s}$.
Each of these edges provides three faces $F_e^1, F_e^2, F_e^3$, namely the faces of
$\sigma_e$ to which $e$ belongs. We say that a face $F$ belongs to $A_s^{(2)}$ if $F \in \sigma_e$ and $F$ admits $e$ as one of its edges.
The assignment $A^{(2)}_{s} = \{\sigma_e\}_{e \notin A^{(1)}_{s}}$ is said admissible\footnote{Whether or not there exists
a systematic way to construct such admissible assignments, is a question that is left open here.} if $F_e^i \not= F_{e'}^j$ for every couple
$(e, e')$ of distinct
edges that are not in $A^{(1)}_{s}$. Eventually we define $A_s = A^{(1)}_s \cup A^{(2)}_s$: a simplex $\sigma$ or a face $F$ belong to $A_s$ if
they belong to one of the assignments $A_s^{(1)}, A_s^{(2)}$. The assignment $A_s$ is said admissible if both $A_s^{(1)}$ and $A_s^{(2)}$ are admissible.

Given an admissible assignment $A_s$, the gauge-fixing term and Faddeev-Popov determinant,
associated to the symmetry (\ref{transf}) read then
\beq \label{gfsym2}
\delta^{A_s}_{GF} = \prod_{F \in A_s} (2\pi) \delta_{s_F, s_F^o}, \qquad \tilde{D}_{FP}^{A_s} = \prod_{v}D^{\sigma_{v}}\prod_{e}D^{\sigma_e}=\frac{\prod_{\sigma \in A_s} \VV_{\sigma}}{\prod_{F \in A_s} \A_F}.
\eeq

\medskip
The gauge fixing described above allows us to fix the value of $l_{e}, s_{F}$ on a subset of edges and faces.
By removing the summation over $s_{F}$, we remove a redundant factor $\delta(\omega_{F}^{\epsilon})$ which naively makes the partition function divergent, this naive divergence being, as we have just shown, the expression of a gauge symmetry acting around Regge solutions.
 It is important to note however that the constraints $\omega_{F}^{\epsilon}=0 \, [2\pi]$ act not only as constraints on the continuous  edge labels,
 but also on the discrete orientation labels $\epsilon$.
 That is, given a  set of $l_{e}$ which describes a flat space geometry, there is only a restricted choice of orientations $\epsilon$
 which allows the realization of this flat space geometry in terms of an oriented triangulation.
Thus if the $\epsilon$ is not chosen appropriately  it is not possible to satisfy the flatness constraint
 $\omega_{F}^{\epsilon}=0\, [2\pi]$ and the corresponding partition function is in fact zero.
 The gauge fixing of $s_{F}$ removes this necessary restriction on the orientations and an additional gauge fixing factor
 acting on the orientation label should be added.
 The additional gauge fixing factor acting on orientations has a natural topological interpretation: in order for the triangulation to
describe a manifold, one should insure that the link around any internal edge as the topology of a $2$-sphere.
As shown in the appendix, this condition can be implemented by demanding that the solid angle $\Omega^{\epsilon}_{e}$ around any edge is equal to $4\pi$,
where $\Omega_{e}^{\epsilon}=\sum_{\sigma \supset e} \epsilon_{\sigma}\Omega_{e}^{\sigma}$ and
$\Omega_{e}^{\sigma}$ is the solid angle of the edge $e$ within the $4$-simplex $\sigma$.
This is realized by adding to the face gauge fixing (\ref{gfsym2}) a term $\prod_{e\in A_{s}} \Theta(\Omega_{e}^{\epsilon})$
and define
\be\label{gepsilon}
 {D}_{FP}^{A_s}\equiv \tilde{D}_{FP}^{A_s}\prod_{e\in A_{s}} \Theta(\Omega_{e}^{\epsilon})
\ee
where $\Theta$ is a characteristic function defined to be constant, with value $1$,
on $4\pi\mathbb{Z}$ and $0$ elsewhere.

%%%%%%%%%%%%%%%%%%%%%%%%%%%%%%%%%%%%%%
\subsection{The gauge-fixed model}
%%%%%%%%%%%%%%%%%%%%%%%%%%%%%%%%%%%%%%

For a given triangulation $\Delta$, the fully gauge-fixed partition function is defined by choosing admissible assignments $A_s$ and $A_l$
and by inserting gauge-fixing terms and Faddeev-Popov determinants (\ref{gfsym1}, \ref{gfsym2}, \ref{gepsilon}) in the integral.
\beq \label{gfmodel}
Z_{\Delta}^{GF} \equiv \frac{1}{(2\pi)^{|F|}} \int  \prod_{e\in\Delta} \mathrm{d}l_e l_e  \prod_{F\in\Delta}\mathcal{A}_{F}
\sum_{\{s_F,\epsilon_{\sigma}\}} \left( \prod_{\sigma}\, \frac{e^{\imath \epsilon_{\sigma} S_{\sigma}(s_F,l_e)}}{\VV_{\s}}\right)
\delta_{GF}^{A_l} \delta_{GF}^{A_s} D_{FP}^{A_l} D_{FP}^{A_s}
\eeq
This analysis completes the definition of the spin foam model (\ref{4dmod}). The next section is devoted to proving that the Feynman integral
(\ref{Feyamp}) is equal to the evaluation of an observable for this model.

%%%%%%%%%%%%%%%%%%%%%%%%%%%%%%%%%%%%%%%%%%%%%%%%%%%%%%%%%%%%%%%%%%%%
\section{Feynman diagrams as spin foam amplitudes} \label{proof}
%%%%%%%%%%%%%%%%%%%%%%%%%%%%%%%%%%%%%%%%%%%%%%%%%%%%%%%%%%%%%%%%%

In this part, we want to show our main statement (\ref{Exp}). To do so, we first establish that the gauge-fixed partition function is
independent of the triangulation - and thus defines an invariant of $4$d-manifold. Then given a Feynman graph $\Gamma$,
we will consider the so-called Feynman graph observable, which breaks part of the gauge symmetry,
promoting gauge degrees of freedom to dynamical degrees of freedom. We properly define the expectation value of this observable
and show that it coincides with the Feynman amplitude (\ref{Feyamp}) associated to the graph $\Gamma$.

%%%%%%%%%%%%%%%%%%%%%%%%%%%%%%%%%%%%%%%%%%%%%%%%%%%%%%%%%%%%%%%%%%%%%%%%%%%
\subsection{Topological invariance}
%%%%%%%%%%%%%%%%%%%%%%%%%%%%%%%%%%%%%%%%%%%%%%%%%%%%%%%%%%%%%%%%%%%%%%%%%%%%%%%%%

Showing that the model does not depend on the triangulation, and depends only on the piecewise linear topology of the underlying manifold,
%\footnote{This establishes that the model defines a smooth manifold invariant.
%In order to show, furthermore, that the model depends only
%on the topology of the manifold that $\Delta$ triangulates, one would also need to prove that it gives a homotopy
%type invariant \cite{Freedman}. We do not investigate such a proof here.} ,
amounts to check that it is invariant under $4$d Pachner moves and under change of the admissible assignments $A_{s}, A_{l}$.
The invariance under change of admissible assignments is a priori insured by the fact that these assignments
arise from the gauge fixing of a gauge symmetry of the theory; this  is the usual BRST symmetry. Indeed the gauge fixing terms and Faddev-Popov determinants are constructed in order to define the right measure of integration
 over the orbit space of configuration modulo gauge transformation.
 It would be nevertheless interesting to give a formal and direct proof of this invariance: we leave this as an open problem and now focus
 on the invariance under Pachner moves.
In this proof we can however show that, whenever a gauge fixing is needed in order to define the Pachner move, the result is independent under the
gauge fixing choice.

We consider six $4$-simplices $\sigma_0, \ldots, \sigma_5$ which triangulate the boundary of a $5$-simplex, $\sigma_i$ being the $4$-simplex
where the vertex $i$ is omitted. The volume of $\sigma_i$ is denoted by $\VV_i$; the action term $S_{\sigma_i}(s_F,l_e)$, defined in (\ref{action}) and
associated to the simplex $\sigma_i$, is simply written $S_i$.

\medskip

We first investigate the behavior of (\ref{gfmodel}) under the move $(3,3): \sigma_1, \sigma_2, \sigma_3
\longrightarrow \sigma_0, \sigma_4, \sigma_5$, which erases the face $\left[045\right]$ and provides the face
$\left[123\right]$. The triangulations arising in both sides of the move do not contain internal faces or edges  therefore no gauge fixing is needed
in this case.
\begin{figure}[h]
\includegraphics[width=9cm]{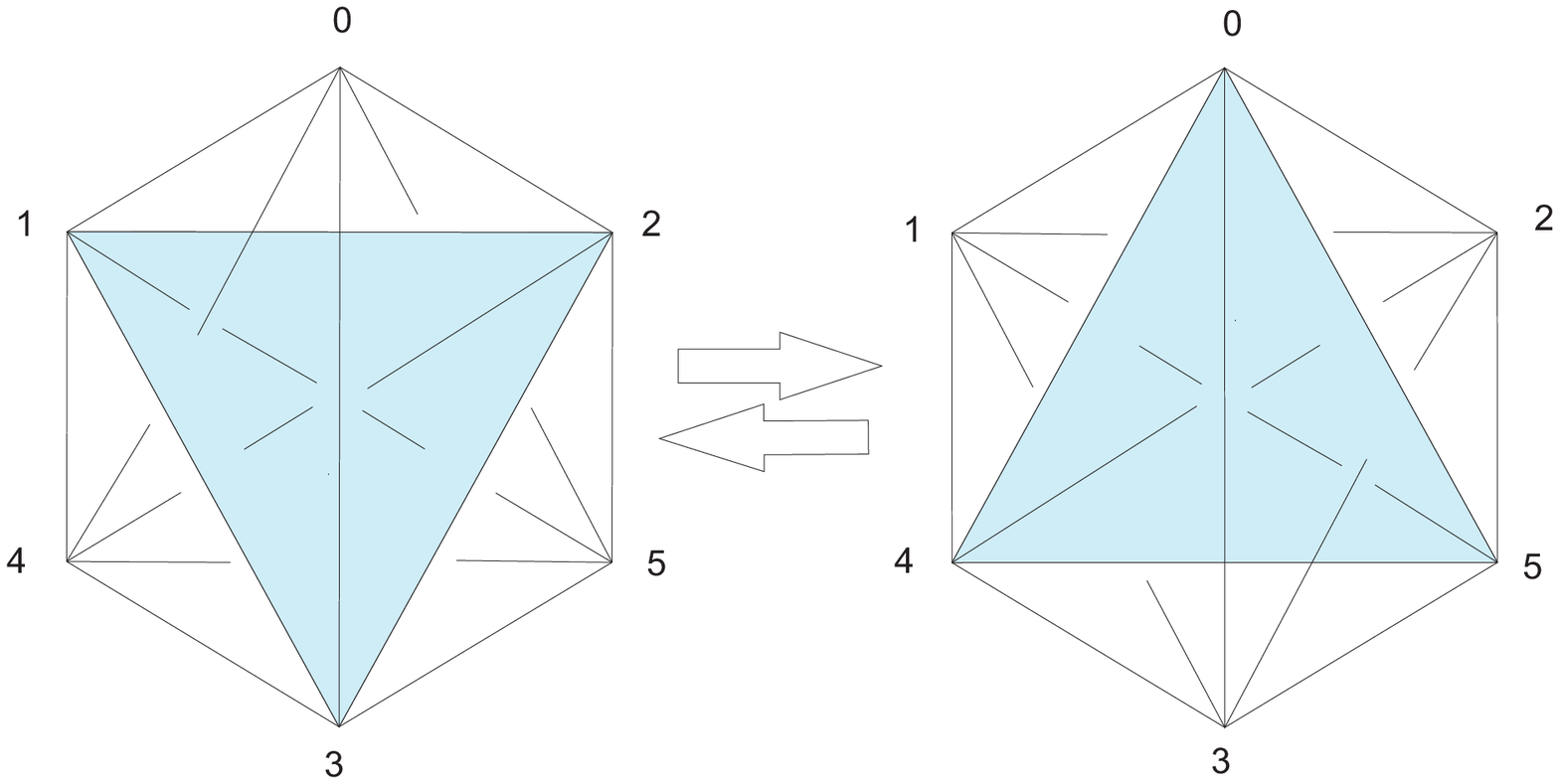}
\label{move1}
\end{figure}

\noindent The invariance of $Z^{GF}_{\Delta}$ under this move is due to the following  $(3,3)$ identity proven in appendix \ref{MovesId}
\beq \label{3,3}
\sum_{\e_1,\e_2,\e_3} \sum_{s_{045}} \A_{045} \frac{e^{\imath \e_1S_1}}{\VV_1}\frac{e^{\imath \e_2S_2}}{\VV_2}\frac{e^{\imath \e_3S_3}}{\VV_3}
= \sum_{\e_0,\e_4,\e_5} \sum_{s_{123}} \A_{123} \frac{e^{\imath \e_0S_0}}{\VV_0}\frac{e^{\imath \e_4S_4}}{\VV_4}\frac{e^{\imath \e_5S_5}}{\VV_5}
\eeq

We then consider the move $(2,4): \sigma_0, \sigma_5 \longleftrightarrow \sigma_1, \sigma_2, \sigma_3, \sigma_4$, which
provides (or erases) the edge $(05)$, as well as the four faces  $\left[05i\right],\, i=1\cdots4$ adjacent to $(05)$.
The complex with 4 $4$-simplices contains one internal edge $(05)$ and four internal faces $\left[05i\right]$ whose labels should be summed over.
Three of the summation over faces variables can be gauge fixed due to the symmetry carried by the edge $(05)$ and acting on faces.
\begin{figure}[h]
\includegraphics[width=10cm]{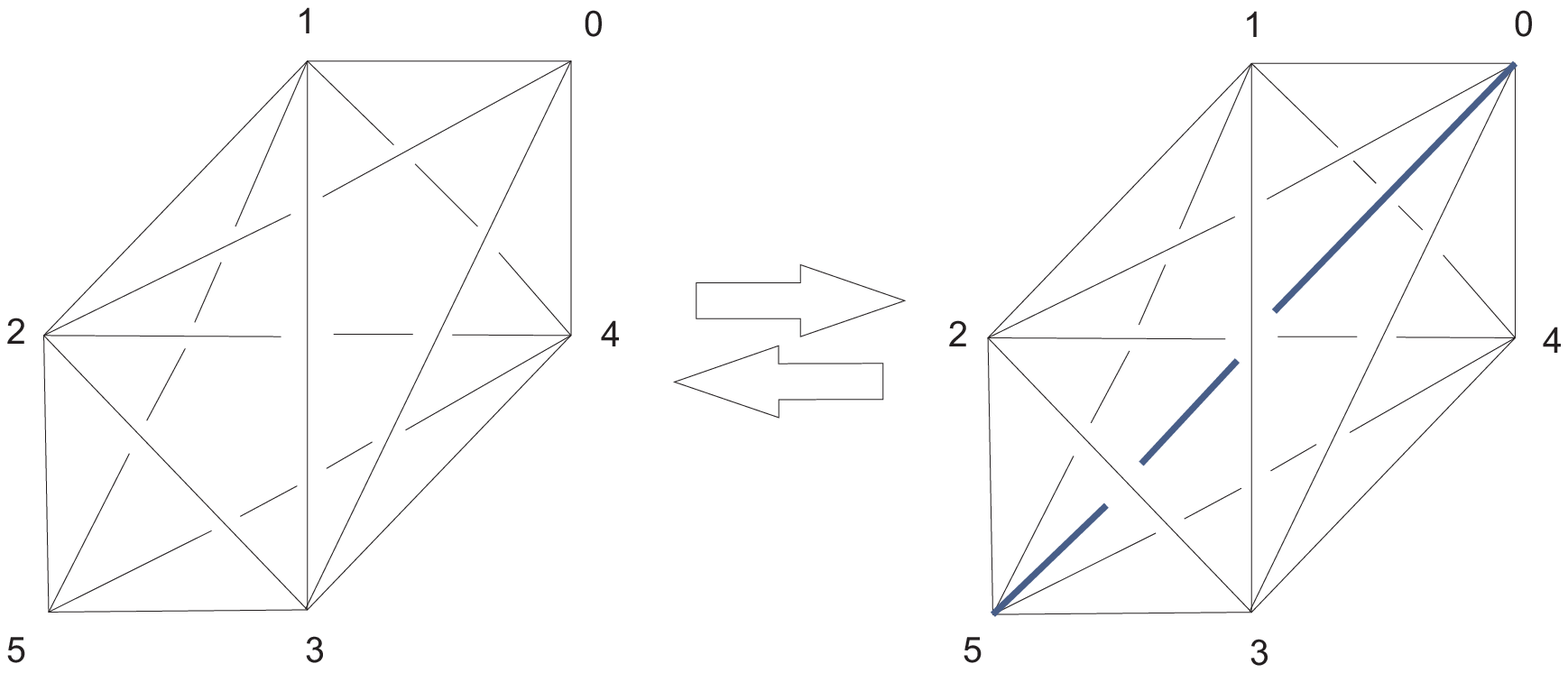}
\label{move2}
\end{figure}

\noindent The invariance under this move follows from the gauge-fixed hexagonal identity
\beq \label{2,4}
\sum_{\epsilon_0,\epsilon_5} \frac{e^{\imath \epsilon_{0} S_{0}}}{\mathcal{V}_0} \frac{e^{\imath \epsilon_{5} S_{5}}}{\mathcal{V}_5}
= \frac{1}{(2\pi)^4} \sum_{\epsilon_i} \int \extd l_{05} l_{05} \sum_{\{s_i\}}
\prod_{i=1}^4  \A_i \prod_{i=1}^4 \frac{e^{\imath \epsilon_{i} S_{i}}}{\mathcal{V}_i} \delta^{(2,4)}_{GF} D_{FP}^{(2,4)}
\eeq
where $s_i \equiv s_{05i}$ and $\A_i \equiv \A_{05i}$ denote label and area of the face $\left[05i\right]$.
The quantities $\delta^{(2,4)}_{GF}$ and $D_{FP}^{(2,4)}$ read
\beq \label{2,4terms}
\delta^{(2,4)}_{GF} = (2\pi)^3 \prod_{i=1}^3 \delta_{s_{i}, s_{i}^o}, \qquad D_{FP}^{(2,4)} = \frac{\VV_4}{\A_1 \A_2 \A_3}\Theta(\Omega_{05}^{\epsilon})
\eeq
where the $s_i^o$ are any fixed values.

These terms can be understood as follows in the case of the move $2 \to 4$: the gauge fixing assignments
of the triangulation containing $2$ simplices need to be extended in order to accommodate for the internal edge $(05)$ added in the move.
According to our prescription, one needs to choose one $4$-simplex ($\sigma_{4}$ say)  and  three faces $\left[05i\right], \, i=1,2,3$
of $\sigma_4$ which are adjacent to $(05)$.
Following (\ref{gfsym2}), the extension of the gauge fixing assignment to $(\sigma_{4}, \left[05i\right])$ implies that
the gauge-fixing term and  Faddev-Popov determinant are then multiplied by (\ref{2,4terms}).
Note that, by symmetry, the identity (\ref{2,4}) is independent of the choice of extension of the gauge fixing assignment.

We eventually examine the change of the model under refinement of the triangulation, namely the move
$(1, 5):\sigma_0 \longleftrightarrow \sigma_1, \sigma_2, \sigma_3, \sigma_4, \sigma_5$, which creates
(resp. erases) the vertex $0$, five edges $(0j), \, j=1\cdots5$ and ten faces $\left[0ij\right],\, i,j=1\cdots5$ meeting at $0$.

\begin{figure}[h]
\includegraphics[width=10cm]{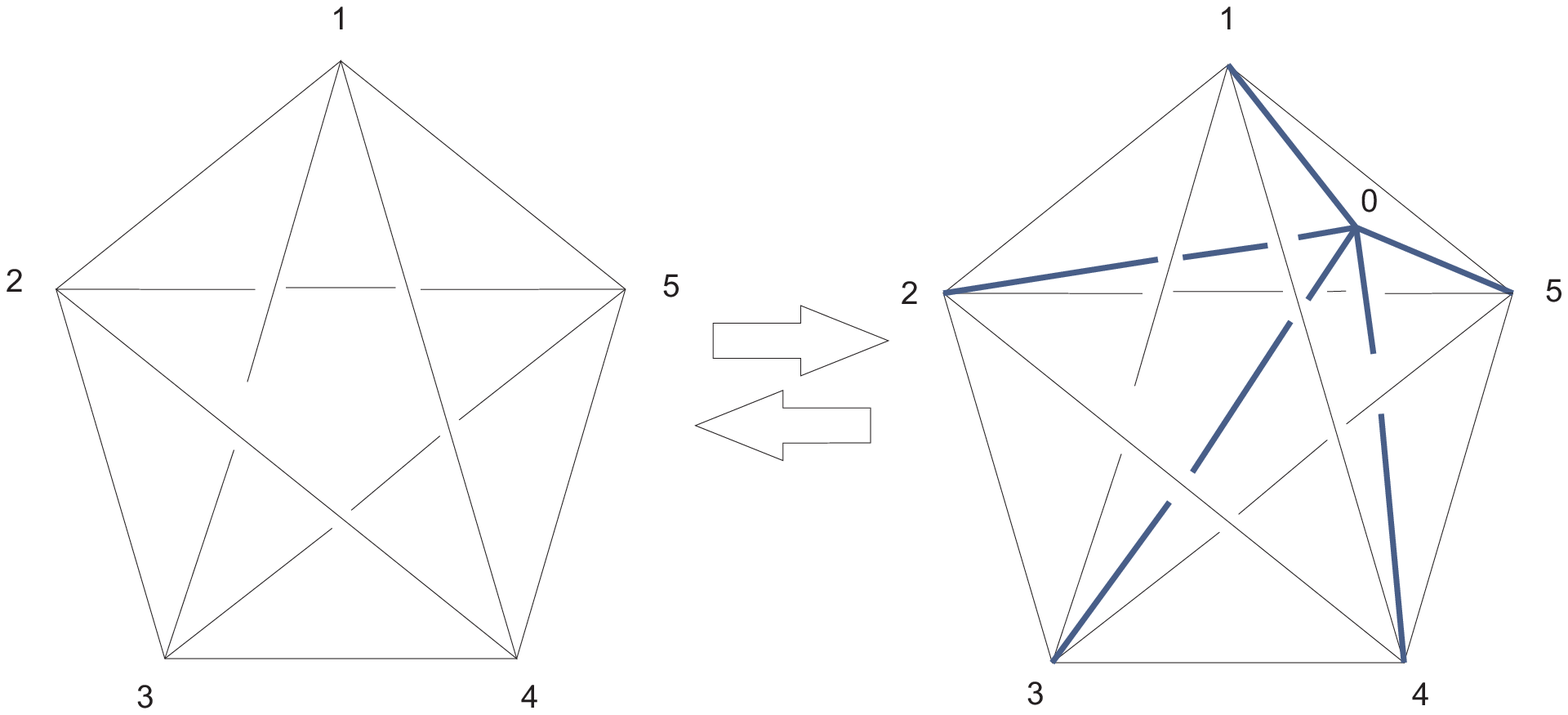}
\label{move3}
\end{figure}

\noindent The invariance under this move is a consequence of the gauge-fixed $(1,5)$ identity
\beq \label{1,5}
\sum_{\e_0}\frac{e^{\imath \epsilon_0 S_0}}{\VV_0} =  \frac{1}{(2\pi)^{10}} \sum_{\e_i} \int \prod_{j=1}^5 \extd l_{0j} l_{0j} \sum_{\{s_{ij}\}}
\prod_{ij} \A_{ij} \prod_{j=1}^{5} \frac{e^{\imath \e_j S_j}}{\VV_j} \delta^{(1,5)}_{GF} D_{FP}^{(1,5)}
\eeq
where $s_{ij} \equiv s_{0ij}$ and $\A_{ij} \equiv \A_{0ij}$ denote label and area of the face $\left[0ij\right]$. The quantities
$\delta^{(1,5)}_{GF}$ and $D_{FP}^{(1,5)}$ read
\beq \label{1,5terms}
\delta^{(1,5)}_{GF} = (2\pi)^{9} \prod_{(ij) \not= (45)} \delta_{s_{ij}, s_{ij}^o} \prod_{i=1}^4 \delta(l_{0i} - l_{0i}^o), \qquad
D_{GF}^{(1,5)} = \left(\frac{\VV_5}{2\prod_{i=1}^4 l_{0i}}\right)
\left(\frac{\VV_5}{\prod_{i,j = 1}^4 \A_{ij}}\right)\left(\frac{\VV_4}{\A_{51}\A_{52}\A_{53}}\right) \prod_{j=1}^5 \Theta(\Omega_{0j}^{\epsilon})
\eeq
where the $l_{0i}^o, s_{ij}^o$ are any fixed values. These terms are those by which
the gauge-fixing terms and Faddeev-Popov determinants (\ref{gfsym1}) and (\ref{gfsym2}) are multiplied when the assignment $A_l$ is
extended to the new vertex $0$, the simplex $\sigma_5$ associated to it, and the four edges $(0i)$ of $\sigma_5$ that meet at $0$; when the assignment
$A^{(1)}_s$ is extended to the new vertex $0$, the simplex $\sigma_5$ associated to it, and the six faces $\left[0ij\right]$ of $\sigma_5$ that share $0$;
and when the assignment $A^{(2)}_s$ is extended to edge $(05)$, the simplex $\sigma_4$ associated to it, and the three faces $\left[05i\right]$
of $\sigma_4$
which are adjacent to $(05)$. The role of the terms (\ref{1,5terms}) is to fix the edge symmetry acting at the new vertex $0$,
as well as the reducible
face symmetry that acts at the new edges $(0i)$ - while taking into account the symmetry of the gauge parameters acting at the vertex $0$.

\medskip

Let us stress that in the $(2,4)$ and $(1,5)$ identities,
once we take into account the additional gauge fixing terms
$\Theta(\Omega_{05}^{\epsilon})$ and $\prod_{j=1}^5 \Theta(\Omega_{0j}^{\epsilon})$,
we can restrict the sum of orientations to be only over $\e_1, \e_2, \e_3$; the values of
$\e_4$ and $\e_5$ are then functions of these orientations and the length $l_{05}$, so that on shell, namely when $\omega_{05}^{\epsilon} = 0$, the other
deficit angles vanish as well.
This is explained  in appendix and  in more detail in \cite{hqg1}.

\medskip

The derivation of these identities is given in appendix \ref{MovesId}. In particular, the keystone of the proof of identity $(2,4)$ is shown to be
the equality of measures (\ref{id4d}). As a consequence, the hexagonal identity (\ref{2,4}) is not only a relation between geometrical quantities,
but it is an equality of measures
which allows to integrate a function of the label $l_{05}$ on the RHS and a function of the other labels and orientations - via the value on shell
$l_{05}^{\e}(l_{ij})$, see section \ref{1} and Appendix - on the LHS.
This remark will be crucial for the demonstration of the statement (\ref{Exp}).

%%%%%%%%%%%%%%%%%%%%%%%%%%%%%%%%%%%%%%%%%%%%%%%%%%
\subsection{Observables and partial gauge-fixing}
%%%%%%%%%%%%%%%%%%%%%%%%%%%%%%%%%%%%%%%%%%%%%%%%%%

We consider a Feynman graph $\Gamma$ and a triangulation $\Delta$ of the $4$-sphere $\SSS^4$ in which $\Gamma$ is embedded.
We define the  Feynman graph observable as a function of the labels $l_e$ living on the edges of $\Gamma$, given by
\beq \label{obs}
\OO(l_e) = \prod_{e \in \Gamma} G^F(l_e)
\eeq
where $G^F$ is the Feynman propagator.
The function $\OO$ in not gauge invariant;  its insertion breaks the symmetry
of the labels $l_e$ which acts at the vertices of the graph $\Gamma$ and thus modifies the gauge-fixing procedure.
The evaluation of this observable is defined to be
\beq \label{eval}
\langle \OO \rangle_{\Delta}  = \frac{1}{(2\pi)^{|F|}} \int_{GF} \prod_{e\in\Delta} \mathrm{d}l_e l_e  \prod_{F\in\Delta}\mathcal{A}_{F}
\sum_{\{s_F,\epsilon_{\sigma}\}} \OO(l_e) \left( \prod_{\sigma}\, \frac{e^{\imath \epsilon_{\sigma} S_{\sigma}(s_F, l_e)}}{\VV_{\s}}\right)
\eeq
where the label $GF$ means that the integral is partially gauge-fixed: namely, the face symmetry is fully gauged fixed as before,
but only the   edge symmetry
acting at the vertices of $\Gamma \setminus \Delta$, is fixed.
In order to fix these  gauge symmetries, we first choose as before an admissible assignment $A_s = A_s^{(1)} \bigcup A_s^{(2)}$.
We then specify  an admissible assignment $A_l = \{\sigma_v\}_{v \notin \Gamma}$, so that a simplex is assigned to every vertex
that does not belong to the graph
$\Gamma$. One can conveniently choose $A_l$ to be the subset of simplices $\sigma_v \in A_s^{(1)}$ associated to the vertices of
$\Delta \setminus \Gamma$.
Inserting the gauge fixing terms (\ref{gfsym1}) and (\ref{gfsym2}) fully fixes the face symmetry and partially fixes the edge symmetry;
the gauge degrees of freedom that are not eliminated couple to the Feynman graph observable, thereby they are promoted to dynamical degrees of freedom.

\medskip

We want to show that the evaluation (\ref{eval}), where the gauge fixing is partially performed, equals the Feynman amplitude $I_{\Gamma}$
associated to
the graph $\Gamma$. Recall that, in (\ref{Feyamp}), the amplitude is expressed as a quantity computed on a triangulation $\Delta_k$ of a $4$-ball $B$,
of a special type: every vertex, edge or face of $\Delta_k$ lie on the boundary. The 1-skeleton dual to $\Delta_k$ is a four-valent tree $T_k$
with open ends. Since $\SSS^4$ can be obtained by gluing two $4$-balls  with reversed orientation along their common boundary $\SSS^{3}$,
we can then construct from $\Delta_{k}$ a triangulation of $\SSS^{4}$ denoted by
 $D\Delta_k \equiv \Delta_k  \sharp_{S^3} \bar{\Delta}_k $.

The reasoning, identical to the $3$d case \cite{hqg1}, is then as follows. Both triangulations $\Delta$ and $D\Delta_k$ of $\SSS^4$ contain
the vertices of the graph $\Gamma$;
they can therefore be constructed out of each other
by a sequence of Pachner moves which do not remove the vertices of $\Gamma$.
Now according to the previous analysis, the quantity $\langle \OO \rangle_{\Delta}$ is invariant under these moves,
provided that the variable $l_e$ living of each edge $e$ erased by a $(2,4)$ move is replaced by its value `on shell' - this is where the remark
of the end of the previous part acquires its importance.
Consequently:
\beq \label{Eval}
\langle \OO \rangle_{\Delta} = \langle \tilde{O}_{\Gamma} \rangle_{D\Delta_k}, \quad \text{with} \quad
\tilde{O}_{\Gamma} \equiv \OO(l_e,l_{e'}^\epsilon({l_e})).
\eeq
The values $l_{e'}^\epsilon(l_e)$ are fully specified by the edge labels of $D\Delta_k$ and the orientations; they are the Euclidean distances,
in any embedding\footnote{Note that, for triangulations of the type of $\Delta_k$, such an embedding always exists.} of $\Delta_k$ in $\RR^4$,
between vertices that are not connected by the edges of the triangulation.

Now with a mechanism explained in \cite{hqg1}, the tree $T_k$ dual to the triangulation $\Delta_k$, once given a root and an orientation, can be used
to define an admissible assignment
$A_s^{(1)} = (\sigma_0, \{\sigma_v\}_{v \notin \sigma_0})$: the simplex $\sigma_0$ is dual to the root of the tree,
and the $\sigma_v$ are defined recursively by following the branches of the tree according to its orientation.
It is then easy to see that every edge of the triangulation belongs to $A_s^{(1)}$; therefore $A_s^{(2)} = \emptyset$ and the face symmetry
is fully fixed by inserting the factors (\ref{gfsym2}) associated to the assignment $A_s \equiv A_s^{(1)}$. The edges, faces and $4$-simplices of
$A_s$ are then the edges, faces and $4$-simplices of $\Delta_k$.
Furthermore, since every vertex of $D\Delta_k$
belongs by construction to the graph $\Gamma$, there is no remaining edge symmetry once the symmetry-breaking observable $\OO$ is inserted
into the partition function. Therefore the gauge fixing is performed by plugging
\beq \label{gfDeltak}
\delta^{A_s}_{GF} = \prod_{F \in \Delta_k} (2\pi) \delta_{s_F, s_F^o}, \qquad D_{GF}^{A_s} =
\frac{\prod_{\sigma \in \Delta_k} \VV_{\sigma}}{\prod_{F \in \Delta_k} \A_F} \prod_{e\in \Delta_k} \Theta(\Omega_{e}^{\epsilon,\epsilon'}).
\eeq
Since $\Delta_k$ has no internal vertex, edge or face, $\Delta_k$ and $D\Delta_k$ possess the same number of vertices, edges and faces,
which all lie on the boundary of the ball $B$, whereas
the number
of $4$-simplices in $D\Delta_k$ is $2k$. Each $4$-simplex in the interior of $B$ has a copy in the exterior of $B$; the two copies share their edges,
and consequently
have the same volume. The orientation of a $4$-simplex $\sigma$ within $B$ and that of its copy
are denoted by $\epsilon_{\sigma}$ and $\epsilon'_{\sigma}$.
Taking into account the gauge fixing terms (\ref{gfDeltak}), the evaluations (\ref{Eval}) read
\beqa
\langle \tilde{O}_{\Gamma} \rangle_{D\Delta_k} &=&
\int \prod_{e\in D\Delta_k} l_e \extd l_e \sum_{\{s_F\}} \prod_{F \in D\Delta_k} \A_F \sum_{\epsilon \in \{\pm\}^{2k}}
\frac{e^{\imath \sum_F s_F\omega^{\e}_F} }{\prod_{\sigma \in D\Delta_k} \VV_{\sigma}}
\OO(l_e,l_{e'}^{\epsilon})\, \delta^{A_s}_{GF} D_{GF}^{A_s} \, \\
&=& \int \prod_{e \in \Delta_k}  l_e \extd l_e  \sum_{\epsilon, \epsilon'\in{\{\pm1\}}^{k}}
\prod _{\sigma \in \Delta_{k}} \frac{1}{\VV_{\sigma}} \OO(l_e,l_{e'}^{\epsilon,\epsilon'})\,  e^{\imath \sum_F s_F^o \omega_F^{\epsilon,\epsilon'}} \prod_{e\in \Delta_k} \Theta(\Omega_{e}^{\epsilon,\epsilon'})
\eeqa
The argument of the exponential, in the latter expression, involves the deficit angles
\beq
\omega_F^{\epsilon, \epsilon'}(l_e) = \sum_{\stackrel{\sigma\in \Delta_k}{ \sigma \supset F}} (\epsilon_{\sigma} + \epsilon'_{\sigma}) \theta_F^{\sigma}
\eeq
We also need to take into account the restriction on the values of $\e, \e'$ imposed by the additional fixing terms (\ref{gepsilon}).
These values are those for which the solid angle at each edge $e$ vanishes modulo $4\pi$: $\Omega_e^{\e, \e'} = 0 \mod 4\pi$ for all $e \in \Delta_k$.
One can now convince oneself that this condition imposes $\epsilon_{\sigma} + \epsilon'_{\sigma} = 0$ for every
$4$-simplex of $\Delta_k$. This finally shows our statement.
\beq
\langle \OO \rangle_{\Delta} =
\int \prod_{e \in \Delta_k}  l_e \extd l_e  \sum_{\epsilon \in{\{\pm1\}}^{k}} \prod _{\sigma \in \Delta_{k}} \frac{1}{\VV_{\sigma}}
\OO(l_e,l_{e'}^{\epsilon}({l_e}))
= I_{\Gamma}
\eeq

\medskip

The conclusion of this analysis is therefore that QFT Feynman amplitudes are obtained by inserting the partially symmetry-breaking observables
(\ref{obs})
into the topological spin foam model (\ref{4dmod}).

%%%%%%%%%%%%%%%%%%%%%%%%%%%%%%%%%%%%%%%%%%%%%%%%%
\subsection{Feynman diagrams on homogeneous spaces}
%%%%%%%%%%%%%%%%%%%%%%%%%%%%%%%%%%%%%%%%%%%%%%%%%

Let us briefly mention how the results established above can directly be extended to spherical and hyperbolical space-times.
The Feynman amplitude of a graph $\Gamma$ embedded in the unit 4-sphere $\SSS^4$ takes the form
\beq \label{Feysph}
I_{\Gamma} = \int_{S^3} \extd u_1 \cdots \extd u_N \prod_{(ij) \in \Gamma} G_m (l_{ij})
\eeq
$\extd u_i$ is the normalized measure on the $3$-sphere. The integrand is a product of propagators, which are functions of the dimensionless spherical
distances $l_{ij} \in \left[0,\pi\right]$ between the vertices, and
invariant under the action of the group $SO(5)$.

Following the strategy used for the flat case, this amplitude is first expressed in terms of the invariant measure
\beq
\sum_{\e \in \{\pm\}^k} \prod_{\Delta_k} \sin l_e \extd l_e \prod_{\sigma \in \Delta_k} \frac{1}{\VV_{\sigma}}
\eeq
$\Delta_k$ is the spherical analogue of the triangulation defined in section \ref{1}; $\VV_{\sigma}$ is the square root of the Gram determinant
$\mbox{det} \left[\cos l_{ij}\right]$ associated to the simplex $\sigma$.

As shown in appendix, all the geometrical identities for flat simplices also hold for spherical simplicies, provided that all `volumes' and `area' are
replaced by the square root of
Gram determinants\footnote{The Gram determinants reduce to Euclidean volumes in the flat limit, see Appendix \ref{MovesId}.}.
The analysis of Feynman graphs on the unit sphere $\mathcal{S}^4$ leads then to the emergence of the spin foam model
\beq \label{sphmod}
Z_{\Delta}= \frac{1}{(2\pi)^{|F|}} \int \prod_{e\in\Delta} \extd l_e \sin l_e  \prod_{F\in\Delta} \A_{F}
\sum_{\{s_F,\epsilon_{\sigma}\}} \left( \prod_{\sigma}\, \frac{e^{\imath \epsilon_{\sigma} S_{\sigma}(s_F, l_e)}}{\VV_{\s}}\right)
\eeq
with an action term  for each $4$-simplex which reads:
\beq \label{actsph}
S_{\sigma} = \sum_F s_F \theta_F^{\sigma}(l_e)
\eeq
where $\theta^{\sigma}_F$ is the spherical interior dihedral angle of the face $F$ in $\sigma$.
The Feynman amplitude $I_{\Gamma}$ is then the expectation value of the partially symmetry-breaking observable
$\OO(l_e) = \prod_{e \in \Gamma} G_m(l_e)$ for the model (\ref{sphmod}), computed on any triangulation $\Delta$ of $\mathcal{S}^4$
which contains $\Gamma$ as a subgraph. Analogous results for hyperbolical space are obtained by working on the hyperboloid and by
replacing all the angles by hyperbolic angles.
The cosmological constant $\Lambda$ is explicitly introduced by re-scaling the spherical lengths $l_e \to L_e = l_e/\sqrt{\Lambda}$.

Notice that the global action of the model no longer admits Regge solutions $(l_e^o, s_F = \alpha A_{F})$ - where $A_{F}$ is the spherical area
of the face $F$ -
unless\footnote{Indeed, the spherical Schl\"afli identity, unlike the flat one, admits a second term proportional to the spherical volume of the simplex.}
$\alpha = 0$;
therefore the analysis of the symmetries has to be done by restricting to fluctuations around degenerate solutions $s_F = 0$. However,
the Regge solutions can be reintroduced by adding to the action (\ref{actsph})
a term $3 \alpha \mbox{Vol}_{\sigma}$ proportional to the spherical volume of the simplex.
It can be checked that all the results mentioned above hold with this modified action as well.

%%%%%%%%%%%%%%%%%%%%%%%%%%%%%%%%%%%%%%%%%%%%%%%%%%%%%%%
\section{Algebraic structure: discussion} \label{alg}
%%%%%%%%%%%%%%%%%%%%%%%%%%%%%%%%%%%%%%%%%%%%%%%%%%%%%%%

In the previous sections we have shown that the state-sum model (\ref{4dmod}) naturally emerges from Feynman amplitudes of ordinary QFT,
and provides dynamics for the background geometry. In the case of 3d Feynman diagrams,
an identical analysis had led us to a dynamical model whose algebraic structure
has been fully dissected \cite{hqg1}: the model turned out to be the spin foam quantization of a BF theory related to $3$d quantum gravity.
Similar results are expected in $4$d. Namely, it should be possible to show that the state-sum (\ref{4dmod})  can be  understood
in terms of an underlying algebraic structure.
Moreover this state sum model should arise as a limit of a model of quantum gravity
 in the regime where usual QFT takes place.
In this part we give some insights into the investigation of the algebraic interpretation of the  model.

%%%%%%%%%%%%%%%%%%%%%%%%%%%%%%%%%%%%%%%%%%
\subsection{A new kind of spin foam model}
%%%%%%%%%%%%%%%%%%%%%%%%%%%%%%%%%%%%%%%%%%%

It is worth introducing, for each simplex $\sigma \in \Delta$, the following 20j-symbol
\beq \label{symb}
\left\{\begin{array}{ccc}
l_{e_1}&\cdots&l_{e_{10}} \\
s_{F_1}&\cdots&s_{F_{10}}
\end{array}
\right\} \equiv \sum_{\epsilon} \frac{e^{\imath \epsilon S_{\sigma}(l_e, s_F)}}{\VV_{\sigma}(l_{e})}=
2 \frac{\cos S_{\sigma}(l_e, s_F)}{\VV_{\sigma}(l_{e})},
\eeq
which depends on ten variables $l_e \in \RR^+$ labeling the edges, and ten variables $s_F \in \mathbb{Z}$ labeling the faces. With our notations
$F_i$ denotes the face opposite to the edge $e_i$ in $\sigma$, while the quantity $S_{\sigma}$ is the action term (\ref{action}) associated to the simplex
$\sigma$. %\be S_{\sigma}(l_e, s_F)= \sum_{F\in \sigma} s_{F} \theta^\sigma_{F}(l_{e}).\ee
We also define the measures
\beq \label{measure}
\int \extd \mu_e \equiv \int l_e\extd l_e \quad \mbox{and} \quad\int \extd \nu_F \equiv \frac{1}{2\pi} \sum_{s_F} \A_F.
\eeq
for each edge $e$ and face $F$ of the triangulation. Note that the face measure $\nu_{F}$ depends, via $\A_{F}$,
on the labels of the three edges that bound $F$.
The model (\ref{4dmod}) takes then the form of a sum over the labels, with the measures (\ref{measure}), of a product of 20j-symbols:
\beq \label{4dmodsymb}
Z_{\Delta} = \int \prod_e \extd \mu_e \prod_F \extd \nu_F \prod_{\sigma}
\left\{\begin{array}{ccc}
l_{e^{\sigma}_1}&\cdots&l_{e^{\sigma}_{10}} \\
s_{F^{\sigma}_1}&\cdots&s_{F^{\sigma}_{10}}
\end{array}
\right\}
\eeq
where $e^{\sigma}_i$ and $F^{\sigma}_i$ label edges and faces of the simplex $\sigma$.

Let us mention two important properties satisfied by the symbols (\ref{symb}).  The first property is the orthogonality relation
\beq
\int \extd \mu_{e_1} \extd \nu_{F_{10}}
\left\{
\begin{array}{ccc} l_{e_1} & \cdots & l_{e_{10}} \\
                   s_{F_1} & \cdots & s_{F_{10}}
\end{array} \right\}
\left\{
\begin{array}{ccc} l_{e_1} & \cdots & l_{e_{10}}' \\
                   s_{F_1}' & \cdots & s_{F_{10}}
\end{array} \right\}
=  2\pi \frac{\delta_{s_{F_1}, s_{F_1}'}}{\A_{F_1}} \frac{\delta(l_{e_{10}} - l_{e_{10}}')}{l_{e_{10}}}
\eeq
which involves two symbols with identical labels except for the face $F_1$ and the edge $e_{10}$:
$s_{F_1} \not= s'_{F_1}$ and $l_{e_{10}} \not= l'_{e_{10}}$. An identical relation holds for every pair $(F, e)$ of face and edge that are not opposite
to each other. The proof of this identity is identical to the derivation of the orthogonality relation for the Poincar\'e 6j-symbol written
in \cite{hqg1} and we do not repeat it here.
The second property is the $(3,3)$ identity (\ref{3,3})
\beq \label{3,3symb}
\int \extd \nu_{\left[123\right]} \prod_{i=4,5,6}
\left\{
\begin{array}{ccc} l_{e^{\sigma_i}_1} & \cdots & l_{e^{\sigma_i}_{10}} \\
                   s_{F^{\sigma_i}_1} & \cdots & s_{F^{\sigma_i}_{10}}
\end{array} \right\}
=
\int \extd \nu_{\left[456\right]} \prod_{j=1,2,3}
\left\{
\begin{array}{ccc} l_{e^{\sigma_j}_1} & \cdots & l_{e^{\sigma_j}_{10}} \\
                   s_{F^{\sigma_j}_1} & \cdots & s_{F^{\sigma_j}_{10}}
\end{array} \right\}
\eeq
between the symbols of six $4$-simplices which triangulate the boundary of a $5$-simplex $\left[1\cdots6\right]$.
The $4$-simplex $\sigma_i$ is the one obtained by dropping the point $i$, and
$\left[ijk\right]$ is the face whose vertices are $i,j,k$. In this relation we have denoted by $l_e^{\sigma_i}, s_{F}^{\sigma_i}$ the labels of edges and faces of
$\sigma_i$ - keeping in mind that, of course, $l_e^{\sigma_i} = l_e^{\sigma_j}$ (resp. $s_F^{\sigma_i}=s_F^{\sigma_j}$) if $\sigma_i$ and $\sigma_j$ share
the edge $e$ (resp. the face $F$). The identity (\ref{3,3symb}) insures, together with
gauge-fixed hexagonal and $(1,5)$ identities,
the topological invariance of $Z_{\Delta}$.

Although constructing models out of symbols, attached to each simplex and function of the coloring, is a common feature of the spin foam approach,
let us emphasize that
the structure revealed in (\ref{4dmodsymb}) is quite unusual. The basic ingredient of a spin foam model is, indeed, a 2-complex whose faces
are colored by representations of a group and edges by intertwiners. If this 2-complex is the 2-skeleton $\mathcal{J}_{\Delta}$ dual to a triangulation,
one can equivalently work with a labeling of the triangular and tetrahedral faces of $\Delta$. Hence, in the usual picture, unlike for our model,
no edge labels are involved. The structure of (\ref{4dmodsymb}) is however reminiscent of that of 2-category state-sum models,
based on representation theory of categorical groups \cite{Mackaay, Barrett, Crane}. We expect, more precisely, our model to be related to the Poincar\'e
2-group representation theory \cite{Baez}.

%%%%%%%%%%%%%%%%%%%%%%%%%%%%%%%%%%%%%%%%%%%%%%%%%%%%%%%%%%%%%%%%%%%%%%%%%%
\subsection{A duality relation}
%%%%%%%%%%%%%%%%%%%%%%%%%%%%%%%%%%%%%%%%%%%%%%%%%%%%%%%%%%%%%%%%%%%%%%%%%%

We would like to mention an intriguing duality relation between
the symbol (\ref{symb}) and the vertex amplitude of
the Barrett-Crane spin-foam model for $4$d quantum gravity \cite{BC}. In this model the vertex amplitude takes the form of a 10j-symbol
which depends on ten variables labeling simple representations of $SO(4)$.
We consider here the Barrett-Crane 10j-symbol associated with the Poincar\'e group instead of $SO(4)$.
The representations are labelled by their mass $m$ and their spin $s$, and the simple representations are those for which $s=0$.
Let $\sigma \equiv \left[12345\right]$ be a $4$-simplex whose edges $(ij)$ carry Poincar\'e simple representations $(m_{ij}, 0)$.
With the technology introduced in \cite{LK}, we know that the corresponding 10j-symbol can be expressed as the Feynman graph evaluation
\beq \label{eval1}
\int_{\RR^4} \prod_{i=1}^5 \extd x_i \prod_{i < j} K_{m_{ij}}(x_i, x_j)
\eeq
where the kernel $K_{m}(x, y)$ is the Hadamard propagator in $\RR^4$, $(\Delta + m^2)K_m = 0, \, K_m(x,x) = 1$, which\footnote{
This kernel is a Bessel function $K_m(|x|) = \frac{2}{m|x|} \mbox{BesselJ}(1,m|x|)$, with
$\mbox{BesselJ}(1,a) \equiv \frac{a}{\pi}\int_{-1}^{+1} \extd u \,\sqrt{1-u^2}\, e^{\imath a u}$.} only
depends on the distance $|x-y|$.
Note however that the quantity (\ref{eval1}), which
corresponds to the contraction, according to the geometry of the 4-simplex, of five Poincar\'e intertwiners attached to the vertices,
is divergent. We get in fact the correct definition of the 10j-symbol
by gauge fixing the
$ISO(4)$ symmetry in the integral and, thus, working with the invariant measure. This invariant measure has been computed in section \ref{1} in terms of
the distances $l_{ij} = |x_i - x_j|$ and the volume of the simplex:
$$ \extd \mu(l_{ij}) = \frac{\prod_{i < j} \extd l_{ij} l_{ij}}{\VV_{\sigma}(l_{ij})}. $$
We see therefore that the Barrett-Crane 10j-symbol can be expressed in terms of a Fourier transform of  the symbols (\ref{symb})
for zero spin
\beq
\{(m_{ij},0)\}_{BC} = \int \prod_{i < j} \extd l_{ij} l_{ij} K_{m_{ij}}(l_{ij})
\left\{\begin{array}{ccc} l_{12} & \cdots & l_{45} \\
                               0 & \cdots & 0
\end{array}
\right\}
\eeq
up to normalization factors.
This relation is reminiscent of the duality relations arising in 3 dimensions and studied in \cite{KarimLP}.
We expect this duality relation to admit a generalization to the case of non trivial spins.

%%%%%%%%%%%%%%%%%%%%%%%%%%%%%%%%%%%%%%
\subsection{Gravity and BF theory}\label{BF}
%%%%%%%%%%%%%%%%%%%%%%%%%%%%%%%%%%%%%%

If we take the view, detailed in the introduction, that the dynamical model (\ref{4dmod}) is the limit $G_N \to 0$ of the quantum gravity
amplitude, then this model is expected to be a spin foam quantization of classical gravity in this sector. In the work \cite{FA},
it is shown that gravity action including an Immirzi parameter can be written as an $SO(5)$ gauge theory
\be
S=\int \left(B^{ij}\wedge( R_{ij}(\omega)- \frac{1}{l^2}e_{i}\wedge e_{j}) + \frac{1}{l}B_{i}\wedge d_{\omega}e^{i}- \frac{\beta}{2} B^{ij}\wedge
B_{ij}-\frac{\beta}{2} B^{i}\wedge B_{i}
-\frac{\alpha}{4} B_{ij}\wedge B_{kl}\epsilon^{ijkl}\right) \label{action1}
\ee
where $e^{i}$ is the frame field, $\omega^{ij}$ the spin connection and $R_{ij}(\omega)$ its curvature,
$B^{ij}, B^{i}$ are $2$-form fields valued respectively in the adjoint and vectorial representation of $SO(4)$.
$l $ is the cosmological  length scale  and $\alpha, \beta$ are dimensionless parameter  expressed in term of
the Newton constant $G_N$, the cosmological constant $\Lambda$ and Immirzi parameter $\gamma$:
\be
\frac{1}{l^2}=\frac{\Lambda}{3},\qquad
\alpha = \frac{G_N\Lambda}{3(1-\gamma^2)},
\qquad \beta = \frac{\gamma G_N\Lambda}{3(1-\gamma^2)}.
\ee
Now when $G_N \to 0 $ , the theory becomes topological. Indeed in this limit, if $\gamma$ is fixed, we have
$\alpha , \beta \to 0$ and therefore the action (\ref{action1}) reduce to a $SO(5)$ BF theory
\be \label{topo}
S=\int
B_{ij}\wedge (R^{ij}(\omega)-\frac{1}{l^2} e^i\wedge e^j)
+ \frac{1}{l} B_{i}\wedge d_{\omega}e^{i}.
\ee
Note that we could consider gravity coupled to matter, described by (\ref{action1}) with an additional matter action
$S_m(e,\phi)$, which depends on the frame fields and on matter fields
denoted collectively by $\phi$: we see that the limit $G_N \to 0$ does not affect the matter sector.
The equations of motion of (\ref{topo}) coming from the variation of the $B$ fields
are
\be
R^{ij}=\frac{\Lambda}{3}e^i\wedge e^j,\quad d_we^i=0.
\ee
The unique solution to these equations is the deSitter, antideSitter or flat space, depending on the value of the cosmological constant.

The topological models (\ref{4dmod}, \ref{sphmod}) are thus expected to be a spin foam quantization of (\ref{topo}), and
the QFT Feynman amplitudes to be related to Wilson lines observables for this theory \cite{FAJ}.

%%%%%%%%%%%%%%%%%%%%%%%%%%%%%%%%%%%%
\section{Conclusion}
%%%%%%%%%%%%%%%%%%%%%%%%%%%%%%%%%%%%%

%Form of the propagator from first principle
%Algebraic interpretation 2-category
%Open Feynman diagram
%possible deformations
%Quantum gravity corrections.

The goal of this paper was to bridge the gap between the language of spin foam models,
which provide a well defined framework to address the dynamical issue of quantum gravity in a background independent way,
and the usual language of quantum field theory. The significance of the results we have obtained is twofold. Firstly,
it gives a background independent perspective to standard field theory, in a way that is in agreement with the spin foam hypothesis.
Indeed, in the formulation we have proposed, background geometry is dynamical and the dynamics is governed by a spin foam model.
Furthermore, this model is revealed to be topological, which  confirms the idea that gravity becomes topological
in the limit $G_N \to 0$. The second interest of our results is that they provide a falsification test for any candidate for
the quantum gravity amplitude;
we indeed claim that it must reduce to the spin foam model (\ref{4dmod}) in a suitable semi-classical limit. This requirement represents strong
constraints on the physically viable proposals for quantum gravity models.

Further investigations regarding the structure of the model are needed, in order to understand the algebraic origin of the quantum weights.
Some indications have been given in the last section, which have to be studied in more detail. The conjecture of an interpretation in terms
of 2-categories needs to be investigated; possible links with the Barrett-Crane model have to be explored; and the connection with the continuum
approach deserves to be precisely established. In our view, dissecting the algebraic structure of the Feynman graph spin foam model is a key step,
hopefully allowing us a new way to propose and study
possible dimensionfull deformations of usual field theory and their relations with quantum gravity models.

Let us emphasize that, throughout the paper, we have assumed that Feynman amplitudes were properly regularized in order to avoid divergences, and hence ignored the issue of renormalization. It would be extremely interesting to investigate how renormalization procedure in QFT can be formulated in the new spin foam context.

We have restricted our analysis to closed Feynman diagrams: it needs to be to extended to the case of open diagrams. The main issue is to reexpress the dependence of Green functions on the positions in terms of boundary spin networks.

Finally, another direction of investigation concerns the nature of the Feynman graph observable itself, which should be eventually understood
as a natural observable from the spin foam point of view.

\bigskip

\bigskip

{\bf Acknowledgments.}
A.B is grateful to the Perimeter Institute for its hospitality during the period
this paper has been written.
This work is partially supported by the Eurodoc program from R\'egion Rh\^one-Aples and by
the Government of Canada Award Program.

\bigskip

\appendix

%%%%%%%%%%%%%%%%%%%%%%%%%%%%%%%%%%%%%%%%%%%%%%%%%%%%%%%%%%
\section{Pachner moves identities} \label{MovesId}
%%%%%%%%%%%%%%%%%%%%%%%%%%%%%%%%%%%%%%%%%%%%%%%%%%%%%%%%%%%%

In this appendix we describe the main steps leading to (\ref{3,3}, \ref{2,4}, \ref{1,5}), which provide the topological invariance
of the model.

\medskip

\textbf{Geometry of the simplex}.
The key ingredient for the proof of the identities
is a $4$d extension of geometrical equalities established in \cite{hqg1}. We consider a spherical $5$-simplex
$(e_0, \cdots e_5)$, and denote by $l_{ij} \in \left[0,\pi\right]$ its lengths and $G = \mbox{det}\left[\cos l_{ij}\right]$ its Gram determinant.
Let $\VV_j$ be the square
root of the Gram determinant associated to the $4$-simplex $\sigma_j$ obtained by dropping the vertex $j$, and $\epsilon_j$ its orientation.
Then derivatives of $G$, with respect to the lengths on one hand, and to deficit angles
on the other hand, are related to the quantities $\VV_j$ in the following way:
\beqa
\label{4d1}
\left.\frac{\pa G}{\pa l_{ij}}\right|_{l_{ij}^{\pm}} &=& \mp 2 \sin l_{ij} \VV_i \VV_j \\
\label{4d2}
\left.\frac{\pa G}{\pa \omega^{\epsilon}_{ijk}}\right|_{l_{ij}^{\pm}} &=& - 2\,\epsilon_l \epsilon_m \epsilon_n \frac{\VV_l \VV_m \VV_n}{\A_{ijk}}
\eeqa
$(ijklmn)$ is any permutation of $(012345)$; $\omega^{\epsilon}_{ijk}$ is the deficit angle of the face $\left[ijkl\right]$
which depends on the lengths
and orientations $\epsilon_l, \epsilon_m, \epsilon_n$; $\A_{lmn}$ is the square root of the Gram matrix associated to the triangle $\left[ijk\right]$.
All lengths are supposed fixed except for $l_{ij}$; and $(l_{ij}^{\pm}, \epsilon_l, \epsilon_m, \epsilon_n)$
is solution of $G=0$ and $\omega_{ijk}^{\epsilon} = 0 \mod 2\pi$, with $l_{ij}^- < l_{ij}^+$.
These equalities can be easily derived by using the technology introduced in the appendix of \cite{hqg1}, and we do not give more details here.

Note that the flat counterpart of this result is recovered in the limit $l_{ij} \rightarrow 0$ while ratios $l_{ij}/l_{kl}$ are kept fixed; in this limit
the Gram matrices reduce to the square of usual Euclidean volumes and we have the following correspondence:
\beq
\VV(e_0,\cdots, e_D) \sim D!V(l_{ij}), \quad \sin{l_{ij}}\sim l_{ij}
\eeq
In the following we work with flat simplices. For any $4$-simplex $\sigma$ with edge lengths $l_{ij}$, and for any face $F$ of $\sigma$,
$\VV_{\sigma}$ denotes $4!$ times the Euclidean volume of $\sigma$
and $\A_{F}$ denotes $2$ times the Euclidean area of $F$. For simplicity, in all the paper, those
quantities are respectively called `volume' of the simplex  and `area' of the face.

\medskip

\textbf{Identity (3,3)}. We consider a length configuration $l^o_{ij}$ for the 5-simplex, which is a solution of $G(l_e)=0$;
and $\epsilon \equiv (\epsilon_j)$ an orientation configuration such that $\omega^{\epsilon}_F = 0$ for every face $F$ of the 5-simplex.
The formula (\ref{4d2}) yields a relation between the functionals $\delta(G)$ and $\delta(\omega^{\epsilon}_{ijk})$, which holds in the neighborhood of
the solution $l^o_{ij}$:
\beq \label{funct}
2 \delta(G) =  \frac{\A_{045}}{\VV_1\VV_2\VV_3} \delta(\omega^{\epsilon}_{045}) = \frac{\A_{123}}{\VV_0\VV_4\VV_5} \delta(\omega^{\epsilon}_{123})
\eeq
where $\epsilon$ denotes $(\epsilon_0, \e_4, \e_5)$ for the first deficit angle and  $(\e_1, \e_2, \e_3)$ in the second one.
The $(3,3)$ identity arises from (\ref{funct}) by using the fact that, for $l_{ij} = l^o_{ij}$,
the actions of the $(3,3)$ move are related by
\beq \label{action_relation}
\sum_{j=0}^5 \epsilon_j S_j^o = \sum_{F \not= \left[054\right], \left[123\right]} s_F \omega_F^{\epsilon} = 0 \mod 2\pi.
\eeq
where the subscript $o$ means that the actions are evaluated for $s_{054} = s_{123} = 0$.
Note that the identity also holds if all orientations are switched $\epsilon_j \to - \epsilon_j$. Taking into account this remark,
we get, from (\ref{funct}) and (\ref{action_relation}):
\beq
\A_{054} \frac{e^{\imath \eta \e_1S^o_1}}{\VV_1}\frac{e^{\imath \eta \e_2S^o_2}}{\VV_2}\frac{e^{\imath \eta \e_3S^o_3}}{\VV_3}
\delta(\omega^{\epsilon}_{054})
= \A_{123} \frac{e^{-\imath \eta \e_0S^o_0}}{\VV_0}\frac{e^{-\imath \eta \e_4S^o_4}}{\VV_4}\frac{e^{-\imath \eta \e_5S^o_5}}{\VV_5}
\delta(\omega^{\epsilon}_{123}) \qquad \forall \eta = \pm 1
\eeq
We then want to sum this equality over $\eta$; now the delta functions act as constraints on orientations
in such a way that summing over $\eta$ amounts to summing
over the values of $\e_1, \e_2, \e_3$
on the left hand side and over the values of $\e_0, \e_4, \e_5$ on the right hand side of the equation.
Eventually, we get the $(3,3)$ identity by extending this analysis to all solutions of $G=0$,
and expanding the $2\pi$-periodic delta functions in series.
\beq
\sum_{\e_1,\e_2,\e_3} \sum_{s_{045}} \A_{045} \frac{e^{\imath \e_1S_1}}{\VV_1}\frac{e^{\imath \e_2S_2}}{\VV_2}\frac{e^{\imath \e_3S_3}}{\VV_3}
= \sum_{\e_0,\e_4,\e_5} \sum_{s_{123}} \A_{123} \frac{e^{\imath \e_0S_0}}{\VV_0}\frac{e^{\imath \e_4S_4}}{\VV_4}\frac{e^{\imath \e_5S_5}}{\VV_5}
\eeq

\textbf{Identity (2,4)}.
Formula (\ref{4d1}) with $(ij) = (05)$ and (\ref{4d2}) with $(ijk) = (123)$ allows one to compute the derivative
$\pa\omega^{\epsilon}_{045}/ \pa l^{\pm}_{05}$, which provides the equalities of measures
\beq
\label{2,4mes}
\frac{\delta(l_{05} - l_{05}^{\pm})} {\VV_0\VV_5} = \frac{l_{05} \A_{045}}{\VV_1\VV_2\VV_3}\delta(\omega^{\eta\epsilon^{\pm}}_{045})
\qquad  \forall \, \eta=\pm1
\eeq
where $\epsilon^{\pm}$ are values of orientations that satisfy $\omega_{045}^{\epsilon^{\pm}}(l_{05}^{\pm})=0$.
By using again a relation, which holds when deficit angles vanish, between the actions
\beq \label{idact}
\sum_{j=0}^5\epsilon_j^{\pm} S_j^o = 0 \mod 2\pi
\eeq
where the subscript $o$ means that $s_{045}=0$, we get
\beq \label{idcle}
\frac{e^{\imath \eta \epsilon_0^{\pm} S^o_0}}{\VV_0} \frac{e^{\imath \eta \epsilon_5^{\pm} S^o_5}}{\VV_5} \delta(l_{05} - l_{05}^{\pm})
= l_{05}\A_{045} \prod_{i=1}^3 \frac{e^{\imath \eta \e_i^{\pm} S_i^o}}{\VV_i} e^{\imath \eta \e_4^{\pm} S_4^o} \delta(\omega^{\eta\e^{\pm}}_{045})
\eeq
Now one can sum over contributions of $\eta$ on the left and the contributions over $\e_1, \e_2, \e_3$ on the right - since $\{\eta \e^{\pm}, \eta=\pm1\}$
are the only solutions of $\omega^{\epsilon}_{045}(l_{05}^{\pm}) = 0$. By also summing over values $l_{05}^{\pm}$ and expanding the delta function,
the equality can be written in the following integral form
\beq
\sum_{\e_0, \e_5} \frac{e^{\imath\epsilon_0 S_0}}{\VV_0} \frac{e^{\imath \epsilon_5 S_5}}{\VV_5} =
\frac{1}{2\pi} \sum_{\e_i} \int \extd l_{05} l_{05} \sum_{s_{045}} \A_{045} \prod_{i=1}^3
\frac{e^{\imath \e_i S_i}}{\VV_i} e^{\imath \e_4 S_4}
\eeq
The RHS of this expression contains a sum over values of three orientations; the value of $\e_4$ is a function of $\e_i$ defined to be such that
(\ref{idact}) holds. We will see below how to restore a summation over the values of an independent variable $\e_4$. The identity $(2,4)$
is then obtained by reorganizing terms in the integrand and inserting the trivial `gauge-fixing' identity
\beq
1 = \frac{1}{(2\pi)^3} \sum_{\{s_{0i}\}} \prod_{i=1}^3  2\pi \delta_{s_{0i}, s_{0i}^o}
\eeq
We get:
\beq \label{2,4app}
\sum_{\epsilon_0,\epsilon_5} \frac{e^{\imath \epsilon_{0} S_{0}}}{\mathcal{V}_0} \frac{e^{\imath \epsilon_{5} S_{5}}}{\mathcal{V}_5}
= \frac{1}{(2\pi)^4} \sum_{\epsilon_i} \int \extd l_{05} l_{05} \sum_{\{s_i\}}
\prod_{i=1}^4  \A_i \prod_{i=1}^4 \frac{e^{\imath \epsilon_{i} S_{i}}}{\mathcal{V}_i} \delta^{(2,4)}_{GF} \tilde{D}_{FP}^{(2,4)}
\eeq
where
$$
\delta^{(2,4)}_{GF} = (2\pi)^3 \prod_{i=1}^3 \delta_{s_{i}, s_{i}^o}  \quad \mbox{and} \quad \tilde{D}_{FP}^{(2,4}) = \frac{\VV_4}{\A_1 \A_2 \A_3}
$$
Notice that we could have inserted in (\ref{idcle}) a function $f(l_{05})$
of the label $l_{05}$; thus, (\ref{2,4app}) has to be understood as an identity of measures allowing one to integrate a function of the free label $l_{05}$
on the RHS and a function $f(l_{05}^{\e_0\e_5}(l_{ij}))$ of the other labels and the orientations on the LHS.

\medskip

\textbf{Identity (1,5)}.
The derivation of $(1,5)$ is similar to the previous one, the starting point being the following equalities of measures
\beq
\frac{\delta(l_{05} - l_{05}^{\pm})} {\VV_0} = \frac{l_{05} \A_{045}}{\VV_1\VV_2\VV_3} \VV_5 \delta(\omega^{\eta\epsilon^{\pm}}_{045})
\qquad  \forall \, \eta=\pm1
\eeq
that arises from (\ref{2,4mes}) - since the volume $\VV_5$ does not depend on $l_{05}$. We easily get
\beq \label{1,5app}
\sum_{\e_0}\frac{e^{\imath \epsilon_0 S_0}}{\VV_0} =  \frac{1}{(2\pi)^{10}} \sum_{\e_i} \int \prod_{j=1}^5 \extd l_{0j} l_{0j} \sum_{\{s_{0ij}\}}
\prod_{ij} \A_{0ij} \prod_{j=1}^{5} \frac{e^{\imath \e_j S_j}}{\VV_j} \delta^{(1,5)}_{GF} \tilde{D}_{FP}^{(1,5)}
\eeq
where
$$
\delta^{(1,5)}_{GF} = (2\pi)^{9} \prod_{(ij) \not= (45)} \delta_{s_{ij}, s_{ij}^o} \prod_{i=1}^4 \delta(l_{0i} - l_{0i}^o),
\quad \mbox{and} \quad \tilde{D}_{FP}^{(1,5)} = \left(\frac{\VV_5}{2\prod_{i=1}^4 l_{0i}}\right)
\left(\frac{\VV_5}{\prod_{i,j = 1}^4 \A_{ij}}\right)\left(\frac{\VV_4}{\A_{51}\A_{52}\A_{53}}\right)
$$
On the RHS of (\ref{1,5app}), the sum over values of three orientations $\e_1, \e_2, \e_3$
is taken over, the values of $\e_4, \e_5$ being those for which a relation similar to (\ref{idact}) is satisfied.
The issue of promoting these two orientations to independent variables is discussed below.

\medskip

\textbf{Orientations}. We want to write $(2,4)$ and $(1,5)$ identities where a summation of all orientations is taken over. To do so, as mentioned
in section (\ref{symmetries}), we add an
additional gauge-fixing term which plays the role of a constraint for the orientations. In order to describe this term,
let us first define the total algebraic solid
angle at an edge $e$. L
et $p_e$ be a point of the edge $e$. Each simplex $\sigma$ to which $e$ belongs is mapped in a copy of $\RR^4$;
a unit $2$d-sphere,
surrounding $p_e$, in the hyperplane $e^{\bot}$ orthogonal to the edge intersects $\sigma$ along a spherical triangle.
The angles of this triangle are the dihedral angles $\theta^{\sigma}_{F}$ of the faces of $\sigma$ meeting at $e$, and its area,
denoted by $\Omega_{\sigma, e}$,
is the solid angle seen at $e$ within $\sigma$. The spherical angles associated to all the $4$-simplices sharing $e$ triangulate a surface, called the link
$L_{e}$ of the edge $e$. The total algebraic angle is then defined to be
\beq
\Omega^{\epsilon}_e(l_e) = \sum_{\sigma \supset e} \Omega_{\sigma,e}^{\epsilon_{\sigma}}
\eeq
where $\Omega^{\e_{\sigma}}_{\sigma, e} = \Omega_{\sigma, e}$ if $\e_{\sigma} = 1$ and $4\pi - \Omega_{\sigma, e}$ if $\e_{\sigma} = -1$. This is an analogue of the vertex
solid angle defined in \cite{hqg1}. If we define the quantity
$\tilde{\omega}^{\epsilon}_F = \sum_{\sigma \supset F} \tilde{\theta}^{\epsilon}_{F, \sigma}$,
where
$\tilde{\theta}^{\epsilon}_{F, \sigma} = \theta_F^{\sigma}$ if $\epsilon_{\sigma} = 1$, and $2\pi - \theta_F^{\sigma}$ if
$\epsilon_{\sigma} = -1$ (note that this quantity equals the deficit angle modulo $2\pi$),  the following relation holds:
\beq \label{GaussBonnet}
\frac{1}{2\pi} \left[\Omega_e^{\epsilon} + \sum_{F \supset e} (2\pi - \tilde{\omega}_F^{\epsilon})\right]  = \chi(L_e)
\eeq
where $\chi(L) = |\sigma| - |\tau| + |F|$ is the Euler characteristic of the surface $L_e$, $|\sigma|, |\tau|$ and $|F|$ being the number of $4$-simplices,
tetrahedra and faces touching the edge $e$, or equivalently the number of triangles, edges and vertices of the triangulation of the link $L_e$.
If $\Delta$ triangulates a manifold, every link $L_e$ is homeomorphic to a $2$-sphere, and therefore $\chi(L_e) = 2$. We also consider the function
$\Theta(x)$ defined to be constant, with value $1$, on $4\pi\mathbb{Z}$, and $0$ elsewhere. The sum over orientations $\e_4$ in $(1, 4)$ identity and
over orientations $\e_4, e_5$ in $(1, 5)$ identity is then restored by inserting the additional gauge-fixing factors
\beq
\Theta(\Omega^{\e}_{05}) \quad \mbox{and} \quad \prod_{j=1}^5 \Theta(\Omega^{\e}_{0j})
\eeq
in (\ref{2,4app}) and (\ref{1,5app}) respectively. One can check that these terms act on shell as Kr\"onecker symbols for the orientations.
They are introduced\footnote{We interpret the restriction on orientations as a gauge fixing.
It is however clear that this restriction deserves a better formulation and a deeper physical understanding.}
to replace the necessary constraints on orientations removed by the gauge fixing of the $s$'s labels.

%%%%%%%%%%%%%%%%%%%%%%%%%%%%%%%%%%%%%%%%%%%%
\section{Computation of determinants} \label{DetComp}
%%%%%%%%%%%%%%%%%%%%%%%%%%%%%%%%%%%%%%%%%%%%%%

In this part, following Korepanov \cite{KorIII}, we present explicit computations of Faddeev-Popov determinants associated
to the gauge-fixing of the reducible face symmetry.
Since the contributions of the simplex of $A_s^{(2)}$ are well understood - they are given by (\ref{detAs2}) - we will focus on the contribution
$\Delta_{FP}^{\sigma_v}$ of a simplex $\sigma_v \in A_s^{(1)}$ assigned to a vertex $v$.  We will show, for a suitable gauge-fixing condition, the relation
\beq \label{value}
\Delta_{FP}^{\sigma_v} = \frac{\VV_{\sigma_v}}{\prod_{F\supset v} \A_F}.
\eeq
$\VV_{\sigma_v}$ is the volume of the simplex and the product is over the area of the six faces sharing $v$.

\medskip

\textbf{Gauge-fixing condition}. We want to fix the symmetry of six labels $s_F$ living on the faces of $\sigma_v$ that touch the vertex $v$, generated by
four $3$-vectors $\vec{\beta}_e$ attached to the edges of $\sigma_v$ meeting at $v$; to do so, we primarily need to fix the symmetry of the gauge parameters
$\vec{\beta}_e$, generated by a $so(4)$-element $\sigma_{\mu\nu}$ attached to $v$. The actions of these symmetries are respectively given by
\beqa \label{symA}
\delta_e s_F &=& - l_e^{-\frac23} \vec{\beta}_e \cdot \frac{\vec{h}_F}{h_F} \\ \label{symB}
\delta_v \vec{\beta}_e &=& l_e^{+\frac23} \sigma( \frac{\vec{l}_e}{l_e} )
\eeqa
All the vectors are defined in a given embedding of the simplex in $\RR^4$.
$\vec{h}_F$ is the vector represented by the height of the point opposite to $e$ within $F$ when its intersection with $e$ is placed at the origin;
$\sigma(\vec{u})$ is the displacement of a vector $\vec{u}$ by the infinitesimal rotation $1_{4\times4} + \sigma_{\mu\nu}$.
The vertices of $\sigma_v$ are denoted by $0,1,2,3,4$, where $0$ is the vertex $v$. Recall that the parameter $\vec{\beta}_i$ attached to the edge
$(0i)$ belongs to the $3$-dimensional space $\vec{l}^{\bot}_i$ orthogonal to the straight line spanned by $(0i)$.
In order to express the gauge-fixing condition,
we define, for each $i=1,2,3$,
a convenient basis $(\vec{x}_i, \vec{y}_i, \vec{z}_i)$ of $\vec{l}^{\bot}_i$ as follows. Given a cyclic permutation $(ijk)$ of $(123)$, we choose
$\vec{x}_i$ in the plane spanned by the triangle $\left[0ij\right]$ and such that $\vec{x}_i\cdot\vec{l}_j > 0$;
next we choose $\vec{y}_i$ in the space spanned by the tetrahedron $\left[0ijk\right]$, orthogonal to the plane $\left[0ij\right]$
and such that
$\vec{y}_i \cdot  \vec{l}_k > 0$; we eventually choose $\vec{z}_i$ to be orthogonal to the tetrahedron $\left[0ijk\right]$ and such that
$\vec{z}_i \cdot \vec{l}_{4} > 0$ - note that the three $\vec{z}_i$ axis coincide. The components of $\vec{\beta}_i$ in these basis are denoted by
$\beta_i^{x_i},  \beta_i^{y_i}, \beta_i^{z_i}$.

The gauge-fixing prescriptions are the following.
We first take advantage of the six independent components of $\sigma_{\mu\nu}$ to fix the values of the three components $\beta_1^{x_1},
\beta_1^{y_1},\beta_1^{z_1}$ of $\vec{\beta}_1$, two components $\beta_2^{y_2}, \beta_1^{z_2}$ of $\vec{\beta}_2$  and one component $\beta_3^{z_3}$ of
$\vec{\beta}_3$. The Faddeev-Popov determinant $\Delta_2$ which arises from the gauge-fixing of this `second-stage' symmetry
is the determinant of the Jacobian
$6\times6$ matrix associated to the function $\beta(\sigma_{\mu\nu})$.
The six remaining parameters, namely $\beta_2^{x_2}, \beta_3^{x_3}, \beta_3^{y_3}$ and the three components of $\vec{\beta}_4$, are then
used to fix the values of six variables $s_{0IJ}, \, I,J =1\cdots4$, labelling the faces that share the vertex $0$. The Faddeev-Popov determinant
$\Delta_1$ which arises from the gauge-fixing of this `first-stage'
symmetry is the determinant of the $6\times6$ Jacobian matrix associated to the function $s_{0IJ}(\beta)$. The Faddeev-Popov determinant
corresponding to the full reducible symmetry reads then
\beq
\Delta_{FP}^{\sigma_v} = \Delta_1 \Delta_2^{-1}
\eeq

\textbf{Value of $\Delta_1$}. The elements of the Jacobian matrix
associated to the function $s_{0IJ}(\beta)$ are derivatives of the
labels $s_{012},s_{013},s_{023}, s_{041},s_{042},s_{043}$ with
respect to the parameters $\beta_2^{x_2}, \beta_3^{x_3},
\beta_3^{y_3}$ and the three components of $\vec{\beta}_4$; these
derivatives can be read out from (\ref{symA}). The matrix is block
triangular, since the variations $\delta s_{012}, \delta s_{013},
\delta s_{023}$ do not depend on $\vec{\beta}_{4}$. Therefore the
desired determinant splits in two factors. We already know how to
compute one of them: it is the Faddeev-Popov determinant which
corresponds to the gauge-fixing, using the vector $\vec{\beta}_{4}$,
of the labels $s_{04i}$ of three faces meeting at $(04)$, and given
by the formula (\ref{detAs2}):
\beq \label{1fact}
\frac{\VV_{\sigma_v}}{\A_{041}\A_{042}\A_{041}} \eeq
The other factor is the determinant of a triangular matrix, since $\delta
s_{012}$ depend neither on $\beta^{x_3}_3$ nor on $\beta_3^{y_3}$,
and $\delta s_{013}$ is independent of $\beta_3^{y_3}$. Therefore it
reduces to the product of the diagonal elements:
$$
\left(\frac{ \pa s_{012}}{\pa \beta_2^{x_2}}\right) \left(\frac{ \pa s_{013}}{\pa \beta_3^{x_3}}\right)
\left(\frac{ \pa s_{023}}{\pa \beta_3^{y_3}}\right)
$$
Now using (\ref{symA}) and our definition of the axis $\vec{x_2}, \vec{x_3}, \vec{y}_3$, one can convince oneself
that\footnote{modulo a sign, which is irrelevant here since we are only interested in the absolute value of the determinants.} the first derivative equals
$l_2^{-\frac23}$ times the cosine of the angle $\alpha_2$ between the faces $\left[012\right]$ and $\left[032\right]$, the second one is simply
$l_3^{-\frac23}$, and the third one equals $l_3^{-\frac23}$ times the sine of the angle $\alpha_3$ between the faces
$\left[013\right]$ and $\left[023\right]$. The second factor therefore reads
\beq \label{2fact}
l_2^{-\frac23} l_3^{-\frac43} \cos\alpha_2 \sin\alpha_3
\eeq
Finally the desired determinant is the product of (\ref{1fact}) and (\ref{2fact}):
\beq \label{Delta1}
\Delta_1 = l_2^{-\frac23} l_3^{-\frac43}  \frac{\VV_{\sigma_v}}{\A_{041}\A_{042}\A_{041}} \cos\alpha_2 \sin\alpha_3
\eeq

\textbf{Value of $\Delta_2$}. In order to compute this second determinant let us first define a convenient basis $(e_1, \cdots e_4)$ of $\RR^4$
in which the matrix elements of the rotation $\sigma_{\mu\nu}$ will be expressed. We choose the orthonormal basis
such that $e_1 = \frac{\vec{l}_{1}}{l_1}$;
$e_2$ belongs to the plane spanned by $\vec{l}_1, \vec{l}_2$, with $e_2 \cdot \vec{l}_2 >0$; $e_3$ belongs to the space spanned by
$(\vec{l}_1, \vec{l}_2, \vec{l}_3)$, with $e_3 \cdot \vec{l}_3 >0$; and $e_4$ orthogonal to
$e_1, e_2, e_3$ and satisfying $e_4 \cdot \vec{l}_4 >0$.
  The matrix elements
of $\sigma_{\mu\nu}$ in this basis are denoted by $\sigma_{IJ}$. The elements of the Jacobian matrix associated to the function $\beta(\sigma_{IJ})$
are derivatives of the components
$\beta_1^{z_1}, \beta_2^{z_2},\beta_3^{z_3}, \beta_1^{x_1}, \beta_1^{y_1}, \beta_2^{y_2}$ with respect to $\sigma_{14}, \sigma_{24},\sigma_{34},
\sigma_{12}, \sigma_{13}, \sigma_{23}$; these derivatives can be read out from (\ref{symB}).
The matrix is bloc diagonal, since for $i=1,2,3$, $\delta \beta_i^{z_i}$ only depend on $\sigma_{14}, \sigma_{24}, \sigma_{34}$, while the three other
variations $\delta \beta_1^{x_1}, \delta \beta_1^{y_1}, \delta \beta_2^{y_2}$ do not depend on $\sigma_{i4}$. Therefore, again,
the determinant splits in two factors.
We already know how to compute the first factor: if we consider the $3$-vector $\vec{\sigma}$ of span$\{e_1, e_2, e_3\} \simeq \RR^3$
whose components are $\sigma_{i4}$,
then (\ref{symB}) yields:
\beq
\delta \left[l_i^{-\frac23} \beta_i^{z_i}\right] = \vec{\sigma}\cdot \frac{\vec{l}_i}{l_i}
\eeq
for $i=1,2,3$. The LHS is thus the variation of the edge lengths under an infinitesimal move, by the vector $\vec{\sigma}$,
of the vertex $0$ within $\RR^3$. The first factor reads then
\cite{hqg1}:
\beq \label{detprov}
\mbox{det}\left[\left(\frac{\delta \beta_i^{z_i}}{\delta \sigma_{j4}}\right)_{ij}\right] = (l_1l_2l_3)^{\frac23}\frac{\VV_{\tau}}{l_1l_2l_3}
\eeq
where $\VV_{\tau}$ is the $3$d-volume of the tetrahedron spanned by $\vec{l}_1,\vec{l}_2,\vec{l}_3$.
Let $\theta_{ij}$ be the angle between the edges $(ik)$ and $(jk)$ within the triangle $\left[123\right]$;
also, let $\alpha_{3}$ be the angle between the faces
$\left[013\right]$ and $\left[023\right]$. By using the relations
$$\VV_{\tau} = l_1 l_2 l_3 \sin\theta_{13} \sin\theta_{23} \sin\alpha_3, \qquad \A_{0ij} = l_il_i \sin\theta_{ij}$$
(\ref{detprov}) can be written as
\beq \label{1fact'}
l_1^{-\frac13} l_2^{-\frac13} l_3^{-\frac43} \A_{013} \A_{023} \sin\alpha_3.
\eeq
The second factor is the determinant of a triangular matrix, since $\delta \beta_1^{x_1}$ only  depends on $\sigma_{12}$ and $\delta \beta_1^{y_1}$ does not
depend on $\sigma_{23}$. It therefore reduces to the product of the diagonal elements:
$$
\left(\frac{\pa\beta_1^{x_1}}{\pa \sigma_{12}}\right)\left(\frac{\pa\beta_1^{y_1}}{\pa \sigma_{13}}\right)\left(\frac{\pa\beta_2^{y_2}}{\pa\sigma_{23}}\right)
$$
Now using (\ref{symB}) and the definition  of the axis $\vec{x}_1, \vec{y}_1, \vec{y}_2$, one can convince oneself that, up to a sign, each of
the first and second derivatives
equals $l_1^{\frac23}$, while the third one equals $l_2^{\frac23}$ times the sine of $\theta_{12}$ times the cosine of the angle $\alpha_2$ between the
faces $\left[012\right]$ and $\left[032\right]$. The second factor therefore reads
\beq \label{2fact'}
l_1^{\frac43} l_2^{\frac23} \sin \theta_{12} \cos \alpha_2 = l_1^{\frac13}l_2^{-\frac13} \A_{012} \cos\alpha_2
\eeq
where we have used, again, the relation $\A_{012} = l_1 l_2 \sin \theta_{12}$ and between `area' and angle in the triangle $\left[123\right]$.
Finally the desired determinant is the product of (\ref{1fact'}) and (\ref{2fact'}):
\beq \label{Delta2}
\Delta_2 = l_2^{-\frac23} l_3^{-\frac43} \A_{012}\A_{013}\A_{023} \cos\alpha_2 \sin\alpha_3
\eeq

\medskip

\textbf{Value of $\Delta^{\sigma_v}_{FP}$}. The Faddeev-Popov determinant is finally obtained by taking the quotient of (\ref{Delta1}) by (\ref{Delta2});
and the statement (\ref{value}) is proved.
\beq \label{Detfinal}
\Delta_{FP}^{\sigma_v} = \frac{\VV_{\sigma_v}}{\prod_{F\supset v} \A_F}
\eeq

%%%%%%%%%%%%%%%%%%%%%%%%%%%%%%%%%%%%%%%%%%%%%%%%%%%%%%%%%%%%%%%%%%%

\end{document}